\documentclass[%
 reprint,
 amsmath,amssymb,
 aps,
notitlepage
]{revtex4-1}

\usepackage{graphicx}
\usepackage{dcolumn}
\usepackage{bm}

\begin{document}

\preprint{APS/123-QED}

\title{Cell-to-cell variability and robustness in S-phase duration from
  genome replication kinetics }

\author{Qing Zhang\,$^{1}$}
  \email{qzhang519@gmail.com}
\author{Federico Bassetti\,$^{2}$}%
 \email{federico.bassetti@unipv.it}
\author{Marco Gherardi\,$^{1,3}$}%
 \email{gocram@gmail.com}
\author{Marco Cosentino Lagomarsino\,$^{1,3,4}$}%
 \email{marco.cosentino-lagomarsino@upmc.fr}
\affiliation{%
 $^{1}$ Sorbonne
  Universit\'es, UPMC Univ Paris 06, UMR 7238, Computational and
  Quantitative Biology, 15 rue de l'\'{E}cole de M\'{e}decine Paris,
  France
and
$^{2}$Dipartimento di Matematica, Universit\`a di Pavia, Pavia, Italy
$^{3}$IFOM, FIRC Institute of Molecular Oncology, Milan, Italy
$^{4}$CNRS, UMR 7238, Paris, France
}%

\date{\today}

\begin{abstract}
  Genome replication, a key process for a cell, relies on stochastic
  initiation by replication origins, causing a variability of
  replication timing from cell to cell.  While stochastic models of
  eukaryotic replication are widely available, the link between the
  key parameters and overall replication timing has not been addressed
  systematically.
  We use a combined analytical and computational approach to calculate how
  positions and strength of many origins lead to a given cell-to-cell
  variability of total duration of the replication of a large region,
  a chromosome or the entire genome.
  Specifically, the total replication timing can be framed as an
  extreme-value problem, since it is due to the last region that
  replicates in each cell.
  Our calculations identify two regimes based on the spread between
  characteristic completion times of all inter-origin regions of a
  genome. For widely different completion times, timing is set by the
  single specific region that is typically the last to replicate in
  all cells.
  Conversely, when the completion time of all regions are comparable,
  an extreme-value estimate shows that the cell-to-cell variability of
  genome replication timing has universal properties. Comparison with
  available data shows that the replication program of three yeast
  species falls in this extreme-value regime.
\end{abstract}

\maketitle


\section{Introduction}

In all living systems, the duration of DNA replication correlates with
key cell-cycle features, and is intimately linked with transcription,
chromatin structure and genome evolution. Dysfunctional replication
kinetics is associated to cancer and found in aging cells.
Eukaryotic organisms rely on multiple discrete origins of replication
along the DNA~\cite{leonard2013dna,gilbert2001making}. These origins
are ``licensed'' during the G1 phase by origin recognition complexes
and MCM helicases, and can initiate replication during S
phase~\cite{bechhoefer2012replication}. Once one origin is activated
(``fires''), a pair of replication forks are assembled and move
bidirectionally. In one cell cycle, one origin already activated or
passively replicated cannot be activated
again~\cite{gilbert2001making}.
Origins have specific firing rates, possibly connected to the
number of bound MCM helicase complexes~\cite{Das2015}, and their
specificity determines the kinetics of replication during S phase, or
``replication program''.

To investigate genomic replication kinetics, DNA copy number can be
measured with microarray or sequencing, as a function of genome
position and time (see, e.g.,
\cite{yang2010modeling,hawkins2013high,agier2013spatiotemporal}). Based
on such high-throughput replication timing data, it is possible to
infer origin positions and the key parameters for a mathematical
description of the replication process (see,
e.g.,~\cite{yang2010modeling,retkute2012mathematical,Baker2012}). Recent
methods also allow to extract the same information from free-cycling
cells~\cite{Gispan2016}.  The mathematical modeling of genome-wide
replication timing data shows that replication kinetics results from
the stochastic mechanism of origin
firing~\cite{bechhoefer2012replication,hawkins2013high}.
In other words, replication timing originates from individual
probabilities of origin firing (and their correlations with genome
state~\cite{Boulos2015,Moindrot2012,pope2014topologically}). In such
models, firing rate of individual origins determine the kinetic
pattern of replication along the chromosomal coordinate, and fork
velocity is typically assumed to be nearly constant along the genome
(in absence of blockage).

Evidence of this stochasticity directly from single cells (which
should give access to relevant correlation patterns) is less abundant.
Importantly, replication timing patterns observed in population
studies can be explained by stochastic origin firing at the
single-cell level~\cite{bianco2012analysis}.
Stochastic activation of origins leads to stochasticity of termination
and cell-to-cell variability of the total duration of replication of a
chromosome, a genomic region, or the whole
S-phase~\cite{hawkins2013high}, with possible repercussions on the
cell cycle. This raises several questions, including how the
individual rates and spatial distribution of origins cooperate to generate
variability in replication timing, the extent of such variability, and
whether it is possible to identify specific regimes or optimization
principles in terms of cell-to-cell variability.
However, such questions have not been systematically addressed in the
available models.

\enlargethispage{-65.1pt}

A series of pioneering studies~\cite{Yang2008,Bechhoefer2007} has used
techniques of extreme-value theory to derive the distribution of
replication times in the particular case where each locus of the
genome is a potential origin of replication, as in the embryonic cells
of \emph{X.~laevis}.  These efforts allowed to clarify the possible
optimization principles underlying the replication kinetics in such
organisms.

Here, we extend this approach to the widely relevant case of discrete
origins with fixed
positions~\cite{Masai2010,mechali2013genetic,gilbert2001making} using
a modeling framework for stochastic replication to investigate the
cell-to-cell variability of the duration of S-phase (or of the
replication of any genomic region such as one chromosome).  We
use analytical calculations based on extreme-value theory and
simulations,
employ experimental data to infer replication parameters and identify
the main features of empirical origin strengths and positions, and
their response to specific changes.

\section{MATERIALS AND METHODS}

\subsection{Model}

 We make use of a one-dimensional nucleation-growth
model~\cite{herrick2002kinetic} of stochastic replication kinetics
with discrete origin locations $x_i$, similar to models available in
the literature~\cite{yang2010modeling,de2010mathematical}.
Activation of origins (firing) is stochastic, and is described as a
non-stationary Poisson process.  The firing rate $A_i(t)$ of the
origin located at $x_i$ is a function of time, $A_i(t)=\lambda_i
t^\gamma\theta(t)$, where $\theta(t)$ is the step function, and
$\lambda_i$ and $\gamma$ are
constants~\cite{yang2010modeling,Yang2008,meilikhov2015scattering}.
We assume that the parameter $\gamma$ and the fork velocity $v$ are
common to all origins, whereas $\lambda_i$, which reflects the
specific strength of each origin, is origin dependent.  The
probability density function (PDF) $f_i(t)$ of the firing time $t$ for
the $i$-th origin, given that the origin fires during that
  replication round, can be obtained as $f_i(t)=A_i(t)
\exp\left(-\int_0^t A_i(\tau) \mathrm{d} \tau\right)$, which gives
\begin{equation}
  f_i(t)=\lambda_i \, t^\gamma \, \theta(t)
       \exp\left(-\lambda_i \frac{t^{\gamma+1}}{\gamma+1}\right).
\end{equation}
When $\gamma>0$, i.e., when the firing rate increases with time,
$f_i(t)$ is a stretched exponential distribution.
When $\gamma=0$, the firing rates are constant and the process is
stationary, so $A_i(t)=\lambda_i$ and $f_i(t)=\lambda_i
\theta(t)e^{-\lambda_i t}$.

Once an origin has fired, replication forks proceed bidirectionally at
constant speed, possibly overriding other origins by passive
replication.  When two forks meet in an inter-origin region,
replication of that region is terminated.
The length of the $i$-th region is defined as $d_i=x_{i+1}-x_i$;
the time when its replication is completed is $T_i$.
The duration of the S phase $T_\mathrm{S}$
is the time needed for all inter-origin regions to be replicated.

\subsection{Fits}

Empirical parameters were inferred through fitting experimental data
from refs.~\cite{hawkins2013high,agier2013spatiotemporal,heichinger2006genome} on DNA copy
number as a function of position and time with the model.  The
positions of replication origins were obtained directly from the
literature and considered
fixed~\cite{hawkins2013high,agier2013spatiotemporal,heichinger2006genome}. The fits are
performed by minimizing the distance between the replication timing
profiles in the model and in the experimental data. This is carried
out by updating the global parameters ($\gamma$ and $v$) and the local
parameters ($\lambda_i$, $i\in\{1,2,...,n\}$) iteratively (Appendix A). The parameters from these fits are presented
in Supplementary Table S1.

\subsection{Simulations}

Our theoretical calculations (described below)
allow to obtain the cell-to-cell variability of $T_\mathrm{S}$ in
special regimes.
{
We compare simulations using the complete information on the locations
and strengths of all origins fitted from the data, with randomized
chromosomes having similar properties. In these randomized chromosomes
we consider the inter-origin distances $d_i$ and the strengths
$\lambda_i$ as independent random variables. } They are drawn from
probability distributions recapitulating their empirical mean and
variability.
More precisely, from the fitted parameters we fix the mean $\left<
  d\right>$ and the standard deviation $\sigma_d$ of the distance, and
the mean $\left<\lambda\right>$ and the standard deviation
$\sigma_\lambda$ of the strength.
The actual distances $d_i$ and strengths $\lambda_i$ are then drawn by
sampling from two gamma distributions
\begin{equation}
\label{eq:gammas}
d_i\sim
\Gamma \left(\frac{\left<d\right>^2}{\sigma_d^2},\frac{\left<d\right>}{\sigma_d^2}\right),
\quad\quad \lambda_i\sim
\Gamma\left(\frac{\left<\lambda\right>^2}{\sigma_\lambda^2},\frac{\left<\lambda\right>}{\sigma_\lambda^2}\right).
\end{equation}
%
The gamma distribution $\Gamma(a,b)$
(parametrized in terms of a shape parameter $a$ and a rate parameter $b$)
has PDF $p(x) \propto x^{a-1}
\exp(-b x)$.  It yields positive values, with mean $a/b$ and variance
$a/b^2$, and it is the maximum-entropy distribution with fixed mean
and fixed mean of the logarithm. We verified that the assumption of a
gamma distribution was in line with empirical data (Fig.~S1).

%
To explore the full range of parameters, we also used stochastic
simulations, which were performed both (i) with the precise origin
locations and strengths fitted from the data, and (ii) with $d_i$ and
$\lambda_i$ drawn randomly as described above.
To avoid the boundary effects of linear chromosomes, we consider
circular chromosomes with $n$ origins, unless specified otherwise
(boundary effects are discussed in the Appendix B and Fig.~S2,
and do not affect our main conclusions.)
To analyze the biologically relevant regimes, we considered
replication kinetics data on different yeast species, from
refs.~\cite{hawkins2013high} and
    \cite{agier2013spatiotemporal}, ran simulations with such
parameters, and compared with the theoretical predictions using the
empirical values for $\sigma_d$, $\sigma_\lambda$ and mean origin
positions and strengths.

\section{BACKGROUND}

\subsection{The S-phase duration is the result of a maximum
operation on the stochastic replication times of inter-origin regions}

\begin{figure}[t]
   \begin{center}
   \includegraphics[width=0.5\textwidth]{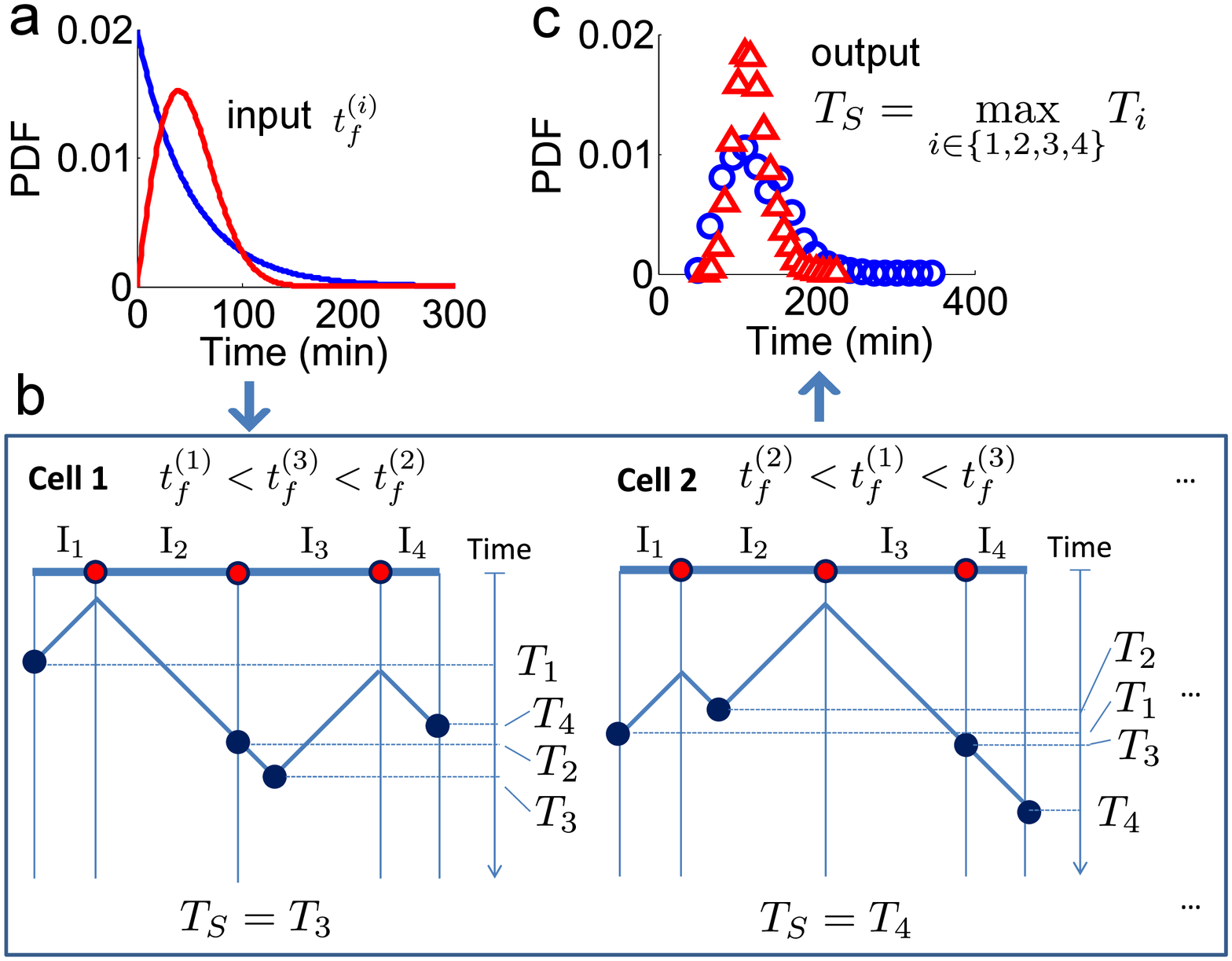}
   \end{center}
   \caption{{\bf The S-phase duration is the maximum between the
       stochastic termination time of all inter-origin regions.}
     The illustration considers replication of one linear chromosome
     with three origins.
     (a) The activation
     of each origin is stochastic, and the firing time $t_f^{(i)}$
     follows a given phenomenological distribution.
     (b) In each cell, each
     origin randomly chooses a firing time from this distribution.
     The last replicated inter-origin region, which may
     be different in different cells, determines the total duration of
     the S phase.
     In the sketch, red circles indicate origins. Dark blue circles
     indicate the latest replicated loci for each inter-origin region.
     Some origins (e.g., the one between $\mathrm{I}_2$ and
     $\mathrm{I}_3$ in cell 1) may be replicated passively, and never
     fire in some realization.  (c) The stochastic model generates a
     distribution of S-phase durations, which expresses the
     cell-to-cell variability. The parameters used in the plots are:
     chromosome length $L=300kb$, fork velocity $v=1 kb/\min$, firing
     exponent $\gamma=0$ (blue line in (a) and blue circles in (c)) or 1
     (red line in (a) and red triangles in (c)),
     origin locations $x_1=50\ kb$, $x_2=150\ kb$ and
     $x_3=250\ kb$, origin strength $\lambda_{1,2,3}=0.02 \min^{-1}$ (for $\gamma=0$)
     or $6.3\times 10^{-4} \min^{-2}$ (for $\gamma=1$).}
  \label{T_SmaxT_i}
\end{figure}

We start by discussing how the stochastic nature of single-origin firing
affects the total replication timing of a chromosome.
Fig.~\ref{T_SmaxT_i}ab illustrates this process.  In each cell, a
chromosome is fully replicated when the last inter-origin region is
complete.  In other words, the last-replicated region sets the
completion time for the whole chromosome.  Consequently, the total
duration is the maximum among the replication times of all
inter-origin regions~\cite{Bechhoefer2007}.
For simplicity, we first consider the case of a genome with only one
chromosome.
The duration of the S phase is therefore
$T_S=\max(T_1,T_2,...,T_n)$ where $n$ is the number of inter-origin
regions.
The stochasticity of the replication time $T_i$ of each inter-origin
region makes the S-phase duration $T_\mathrm{S}$ itself stochastic,
thus giving rise to cell-to-cell variability, which can be estimated
by the model (Fig.~\ref{T_SmaxT_i}c). In the case of multiple
chromosomes, the same reasoning applies to the last-replicated
inter-origin region over all chromosomes.

\section{RESULTS}

\subsection{A theoretical calculation reveals the existence of two
  distinct regimes for the replication program}

  \begin{figure}[t]
  \begin{center}
  \includegraphics[width=0.5\textwidth]{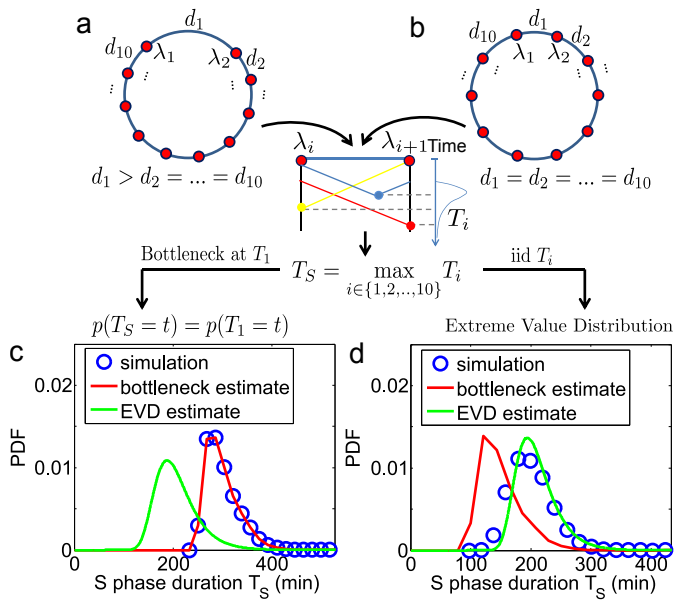}
  \end{center}
  \caption{{\bf Analytical estimates indicate the existence of two
      replication regimes.}  (a) If a single ``bottleneck''
    inter-origin region (labelled by the index 1 in panels a and b) is
    typically the last to complete replication, $T_\mathrm{S}$
    will be typically equal to $T_1$ (inter-origin distances in the
    example are $d_i=167\ kb$ for all origins except $d_1=500kb$).
    (b) If the replication times of all inter-origin regions are
    comparable, and they are considered independent and
    identically-distributed (iid) random variables, the distribution
    of $T_\mathrm{S}$ can be obtained by extreme-value-distribution
    (EVD) theory (inter-origin distances are $d_i=200\ kb$).
    Simulations of the model (blue circles), when one inter-origin
    distance is much larger than the others (c), and when all
    inter-origin distances and strengths are comparable (d), agree with
    the corresponding analytical calculations (red and green curves).
    (Origin number $n=10$ origins, fork velocity $v=1\ kb/\min$,
    origin strength $\lambda_i=0.02 \min^{-1}$.)  }
	\label{twoRegimes}
\end{figure}

It is possible to estimate the distribution of $T_\mathrm{S}$
analytically, starting from the distribution of $T_i$.
Two distinct limit-case scenarios can be distinguished.  In the first
scenario, a specific inter-origin region $r$ is typically the slowest
to complete replication and thus represents a ``replication
bottleneck''.  In this case, $T_\mathrm{S}$ is dominated by $T_r$,
meaning that $T_\mathrm{S}\approx T_r$.
$T_r$ is identified as the one which is largest on average.
Fig.~\ref{twoRegimes}a shows an example chromosome with 10 origins
with the same strength, where one inter-origin distance ($d_1$) is
much larger than the others.
Owing to this disparity, $T_1$ is very likely  the maximum among
all $T_i$, and is therefore the region determining $T_\mathrm{S}$.
In this scenario, which we term ``bottleneck estimate'', the
distribution of $T_\mathrm{S}$ will be approximately the same as that
of the bottleneck $T_r$ (Fig.~\ref{twoRegimes}c).

In the second scenario, each inter-origin region has a similar
probability to be the latest to complete replication.
In this case, every inter-origin region contributes to the distribution
of $T_\mathrm{S}$.  Since $T_S=\max(T_1,T_2,\ldots,T_n)$, we apply the
well-known Fisher-Tippett-Gnedenko theorem
\cite{GnedenkoKolmogorov:1954,Zolotarev:BOOK}, which is a general
result on extreme-value distributions (EVD).
In order to use this theorem, we make the following two assumptions:
(i) $T_1, T_2, \ldots, T_n$ are statistically independent, i.e.,
each inter-origin replication time is an independent random variable,
incorporating the essential information about origin variability and rates;
(ii) $T_i$ follows a stretched-exponential distribution,
independent of $i$, i.e.
\begin{equation}
\label{Ti}
p(T_i<t)=1-e^{-\alpha(t-t_0)^\beta},
\end{equation}
when $t>t_0$, while $p(T_i<t)=0$ when $t\leqslant t_0$. The (positive)
parameters $\alpha$, $\beta$ and $t_0$, effectively describe the
consequences of the model parameters $v$, $\gamma$, inter-origin
distances ($d_1, d_2, ..., d_n$) and origin strengths
($\lambda_1,\lambda_2,...,\lambda_n$) on completion timing of
inter-origin regions (see below and Appendix D), and can be
obtained by fitting the distribution of replication time for a typical
inter-origin region (obtained from simulations) with Eq.~\ref{Ti}.

Our fits show that Eq.~\ref{Ti} is a remarkably good phenomenological
approximation of the distribution of $T_i$ (see Appendix C and
Fig.~S3), thus justifying assumption (ii) above.
Note that the fitted stretched exponential form also incorporates
effectively the coupling existing between different inter-origin
regions.
Indeed, neighboring regions are correlated since they use a pair of
replication forks stemming from their common origin.  Moreover, even
distant inter-origin regions can share the same fork if they are
passively replicated.
In order to justify the assumption (i), we tested the effect of the
correlation between different regions, by sampling $T_1, T_2, \ldots,
T_n$ from the distribution in Eq.~\ref{Ti} independently and then
taking their maximum $T_\mathrm{S}^*$. We verified that the difference
between the distribution of $T_\mathrm{S}^*$ and that of
$T_\mathrm{S}$ obtained from simulation (where the correlations are
present) is small.  Therefore, the effect of {these relatively
  short-ranged correlations can be, to a first approximation,
  neglected at the scale of the chromosomes and of the genome, and
  described by the effective stretched-exponential form} (see
Fig.~S4).

Based on these assumptions, we can use the Fisher-Tippett-Gnedenko
theorem and derive the following cumulative distribution function for
$T_\mathrm{S}$ as a function of the number of origins $n$
and the parameters $\alpha$, $\beta$ and $t_0$ (the
calculation is detailed in the Appendix D):
\begin{equation}
  P(T_\mathrm{S}\leq t)\approx\exp\left\{-\exp\left[\beta\log
      n\left(1-(\alpha/\log n)^{1/\beta}(t-t_0)\right)\right]\right\}.
\label{Ts}
\end{equation}
 Eq.~\ref{Ts} gives a direct estimate of
the distribution of the S-phase duration in this second scenario,
which we term ``extreme-value'' or ``EVD'' regime.
The resulting
distribution is universal, since it does not depend on the detailed
positions and rates of the origins, and depends in a simple way on the
parameters $\alpha$, $\beta$, $t_0$ and $n$.
Although the extreme-value estimate should apply to the case of
  large $n$, the approximation Eq.~\ref{Ts} holds to a satisfactory
  extent also for realistic values, when $n$ is order 10 (see
  Supplementary Fig. S12).
We also derived approximate analytical expressions for $\alpha$,
$\beta$ and $t_0$ as functions of the parameters $v$, $\gamma$, for a
``typical'' region characterized by $\left<\lambda\right>$ and
$\left<d\right>$ under the assumption of negligible interference from
non-neighbour origins (see Appendix D).

The procedure by which we apply Eqs.~\ref{Ti} and~\ref{Ts} is the
following. Given inter-origin distances and origins strengths assigned
arbitrarily or inferred from empirical data, the simulation of the
replication of a chromosome gives the distribution of $T_i$ and
$T_\mathrm{S}$. A fit of the distribution of $T_i$ from simulation
using Eq.~\ref{Ti} gives the parameters $\alpha$, $\beta$ and $t_0$. Finally,
the EVD estimate for the distribution of $T_\mathrm{S}$, can be
obtained from Eq.~\ref{Ts}, and compared with the distribution of
$T_\mathrm{S}$ form simulations.
This procedure can be seen as a variant of the method introduced
in refs.~\cite{Yang2008,Bechhoefer2007} applicable to the case of discrete
origins (see Discussion).

Fig.~\ref{twoRegimes}b shows one example where one circular chromosome
has 10 origins with identical strengths and identical inter-origin
distances. The estimated distribution of S-phase duration from
Eq.~\ref{Ts} is well-matched with the simulated one (Fig.~\ref{twoRegimes}d).
Fig.~\ref{twoRegimes} also shows how the bottleneck estimate works for
the opposite scenario, and compares simulations with both estimates in
the two different regimes.
Similar to Fig.~\ref{twoRegimes}, Supplementary Fig.~S5 shows the
existence of the two regimes in presence of a single origin affecting
the two neighboring inter-origin regions.  In the bottleneck regime,
these two regions replicate much later than the others, because their
common origin is much weaker than the other origins; the S-phase
duration is then dominated by their replication time. This case also
illustrates how the bottleneck regime may not be limited to a single
inter-origin region.
Finally, Supplementary Fig.~S6 shows the distribution of the
inter-origin completion times $T_i$ in the cases presented in
Fig.~\ref{twoRegimes} and Supplementary Fig.~S5. This analysis
illustrates how extra peaks in the right tail of $T_i$ distribution
relate to the failure of the extreme-value estimate for the
distribution of S-phase duration. These examples indicate that, as
expected, the presence of outliers in the values of $T_i$
(exceedingly slowly-replicating regions) is responsible for the
onset of the bottleneck behavior.

\subsection{The extreme-value regime is robust to perturbations
  increasing the replication timing of a local region}

Origin number, origin strengths and inter-origin distances can be
perturbed due to genetic change (DNA mutation or recombination), over
evolution, and due to epigenetic effects such as binding of specific
agents.  We can compare the robustness of the two regimes identified
above to perturbations of these parameters.  We consider in particular
the elongation of a single inter-origin distance $d_i \mapsto d_i +
\delta_d$ (similar results to those reported below are obtained
  for a perturbation affecting the strength of a single origin, see
  Supplementary Fig.~S7).
In such case, the change of $T_i$ is approximately equal to $\delta_d/2v$.
In the bottleneck regime, if the perturbed inter-origin region is
the slowest-replicating one, $\left<T_\mathrm{S}\right>$
increases linearly with $\delta_d$ with slope $1/2v$, and the
distribution of $T_\mathrm{S}$ shifts by a delay $\delta_d/2v$
(Fig.~\ref{perturbation}a).
In the extreme-value regime, instead, there is no single bottleneck
inter-origin region, and the change of
$T_\mathrm{S}$ with the perturbation turns out to be much smaller than
$\delta_d/2v$ (Fig.~\ref{perturbation}b).
Notice that in both regimes the variability of the S-phase duration
around its average is not affected sensibly (insets of
Fig.~\ref{perturbation}).

In summary, the bottleneck
region is ``sensitive''
to the specific perturbations considered, since termination of
replication is highly dependent on a single inter-origin region, while
the EVD regime is ``robust'', as the effect of small local
perturbations can be absorbed by passive replication from nearby
  origins~\cite{hawkins2013high}.

\begin{figure}
  \begin{center}
  \includegraphics[width=0.5\textwidth]{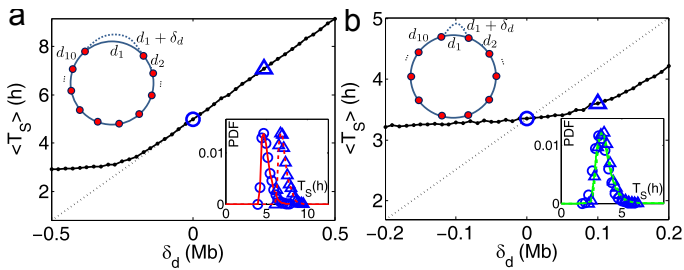}
  \end{center}
  \caption{{\bf Effects of perturbations of a single inter-origin
      region on S-phase duration.}  (a) The bottleneck inter-origin
    region of the chromosome shown in Fig.~\ref{twoRegimes}a is
    perturbed by increasing its length by $\delta_d$ (i.e.,
    $d_1\rightarrow d_1+\delta_d$).  The black solid line with points
    is the average S-phase duration, which increases linearly with
    $\delta_d$.  The black dotted line, with slope $1/(2v)$, is a
    guide to the eye. The inset shows that the perturbation shifts the
    distribution of $T_\mathrm{S}$ by $\delta_d/2v$ (circles are
    simulations for the unperturbed chromosome, and triangles
    correspond to $\delta_d=d_1/2$; the two curves are the analytical
    estimates in the bottleneck regime).  (b) The same perturbation as
    in (a) is performed on an inter-origin region of the chromosome
    shown in Fig.~\ref{twoRegimes}b, which lies in the EVD regime.
    Symbols are as in (a).  The distribution of $T_\mathrm{S}$ is
    robust to this perturbation.  }
\label{perturbation}
\end{figure}

\subsection{Diversity between completion times of inter-origin
  regions sets the regime of the replication program}

The cases discussed above (Fig.~\ref{twoRegimes}) recapitulate the
expected behavior  in case of high versus small variability of
the typical completion time of different inter-origin regions.
One can expect that if the variability of the inter-origin distances
is large, or origin strengths are heterogenous, it will be more likely
to produce a bottleneck region, which in turn will trivially affect
replication timing.
Conversely, the replication program will be in the extreme-value regime
if the completion times of all regions are comparable.
In order to show this, we tested systematically how average and
variability of $T_\mathrm{S}$ change with the variability of
inter-origin distances and origin strengths in randomly generated
genomes.  In this analysis, origin spacings and strengths are
  assigned according to the prescribed probability distributions shown
  in Eq.~\ref{eq:gammas}, with varying parameters (see the Methods for a
  precise description of how chromosomes are generated).

\begin{figure}
\begin{center}
  \includegraphics[width=0.5\textwidth]{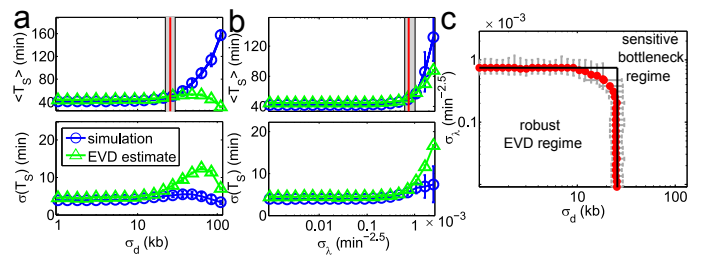}
  \end{center}
  \caption{{\bf The variabilities of the inter-origin distances,
      $\sigma_d$, and of the firing strengths, $\sigma_\lambda$, set
      the replication regime.}
  (a,b) Average S phase duration (top panels) and its standard
  deviation (bottom panel)
  as functions of $\sigma_d$ (panel a) or $\sigma_\lambda$ (panel b),
  obtained by simulations of the model (blue circles and lines)
  and by the EVD estimate (green triangles and lines).
  50 samples of inter-origin distances and origin strengths are
  chosen according to the distributions in
  Eq.~(\ref{eq:gammas}). Red lines indicate the transition points where the
  simulated $\left<\left<T_S\right>\right>$ is $20\%$ larger than at
  $\sigma_d=0$ and $\sigma_{\lambda}=0$.
  The border lines of the grey area show the transition points for
  $\left<\left<T_S\right>\right>+\sigma(\left<T_S\right>)$ and
  $\left<\left<T_S\right>\right>-\sigma(\left<T_S\right>)$
  respectively.
  (c) Phase diagram separating the EVD and bottleneck regimes.  Red
  transition points with error bars (obtained with the method shown in
  (a) and (b)) form an approximate rectangle phase boundary.
  Parameters: fork velocity $v= 1.81~kb/\min$, origin number $n=20$,
  $\gamma=1.5$,
  $\left<d\right>= 28.13 kb$, $\left<\lambda\right>= 6.17\times10^{-4}
  \min^{-2.5}$, $\sigma_{\lambda}= 0$ (a) and
  $\sigma_d = 0$ (b).}
    \label{rectangle}
\end{figure}

Fig.~\ref{rectangle} shows the results.  Importantly, we find that
{the regimes defined above as extreme cases apply for most
  parameter sets, and there is only a small region of the parameters
  where we find intermediate cases}.
Specifically, two parameters, the standard deviations $\sigma_d$ and
$\sigma_\lambda$, of the inter-origins distances and the origin
strengths respectively, are sufficient to characterize the system.
Fig.~\ref{rectangle}a indicates that as long as $\sigma_d$ is smaller
than a threshold (around 30 kb), the average
$\left<T_\mathrm{S}\right>$ and the standard deviation
$\sigma(T_\mathrm{S})$ of the replication time are approximately
constant.  In this regime, the extreme-value estimate matches well the
simulation results.  When $\sigma_d$ exceeds the threshold, the
average of $T_\mathrm{S}$ increases and its standard deviation
decreases with large fluctuations.  In this other regime, both
$\left<T_\mathrm{S}\right>$ and $\sigma(T_\mathrm{S})$ deviate from
the EVD estimate.
Fig.~\ref{rectangle}b shows that varying $\sigma_\lambda$ at fixed
origin positions produces a similar behavior (although with smaller
deviations from the EVD estimates).
%

This analysis shows an emergent dichotomy between these two regimes,
which depends on the distribution of $T_i$ (i.e. both inter-origin
distances and origin firing rates). In principle, more complex
situations where e.g. a subset of many comparably ``slow''
inter-origin regions dominates S-phase timing is possible, but this
situation is very rare (and negligible) if origin rates and positions
are generated with the criteria used here (given by
Eq.~\ref{eq:gammas}).
\emph{De facto}, under these prescriptions, motivated by empirical
properties of origin positions and strengths, only the two regimes
defined above as extreme cases were observable.  For example, one
  can imagine a situation where each chromosome are, separately, in
  the EVD regime, but the replication of one of the chromosomes takes
  considerably longer than the others on average, which may lead the
  S-phase duration to be in the bottleneck regime. However, we find
  that this situation is essentially never found if origin rates and
  positions have empirically relevant values (i.e. for all
  realizations with empirical means and variances of inter-origin
  distances and origin firing rates).
Qualitatively, this will always be the case if the distribution of
$T_i$ shows a single mode, and there are very few, or just one
  exceptional late-replicating region.

This behavior suggests to define ``critical values'' of $\sigma_d$
and $\sigma_\lambda$, separating the extreme-value regime from the
bottleneck regime, as follows.
We define the $\sigma^\mathrm{c}_d$, at fixed $\sigma_\lambda$, as the
value of $\sigma_d$ at which $\left<T_S\right>$ (possibly averaged
over many samples of the origin configuration too, denoted
$\left<\left<T_S\right>\right>$) is $20\%$ larger than at $\sigma_d=0$
and $\sigma_\lambda=0$.  The results presented here do not depend
appreciably on this threshold and do not change much if we define
$\sigma^\mathrm{c}_d$ as the value of $\sigma_d$ at which
$\left<T_S\right>$ is $20\%$ off the prediction of the EVD theory.
The same definition holds for $\sigma^\mathrm{c}_\lambda$ at fixed
$\sigma_d$.
Surprisingly, $\sigma^\mathrm{c}_d$ turns out to be independent of
$\sigma_\lambda$, and $\sigma^\mathrm{c}_\lambda$ independent of
$\sigma_d$.
The resulting ``phase diagram'', shown in
Fig.~\ref{rectangle}c, separates the space of parameters into an approximately
rectangular region where the EVD estimate is precise, and an
outer region where heterogeneities dominate, which is identified with
the bottleneck regime.

We can give a simple argument for why this phase diagram is approximately
rectangle-shaped.  Intuitively, a large $\sigma_d$ increases the
probability of extracting a very large value for $d$, and a large
$\sigma_\lambda$ increases the probability of extracting a very small
$\lambda$.  In a realization of a randomized chromosome, such rare
events may generate an extremely slow-replicating region acting as the
bottleneck.
Clearly, drawing an extreme value for only one of the two variables is
sufficient to generate the bottleneck region, giving rise to the two
sides of the rectangle.
For values of the variances of both variables that are below the
individual thresholds, drawing a large $d$ and small $\lambda$
jointly makes the upper-right region of the rectangle rounded.
However, such joint extreme draws in the same
inter-origin region are very rare, because the two variables are drawn
independently, so the rounded upper-right corner is very small, as
visible in Fig.~\ref{rectangle}c.

\subsection{The yeast replication program is just inside the EVD
  regime and likely under selection for short S-phase duration}

The results of the previous section indicate that the standard
deviations of the origin distances and of the strengths are the most
relevant parameters determining the regime of the distribution of the
S-phase duration across cells.
We inferred the parameters from replication timing data of the yeasts
\emph{S.~cerevisiae} (ref.~\cite{hawkins2013high}),
\emph{L.~kluyveri} (ref.~\cite{agier2013spatiotemporal}) and \emph{S. pombe} (ref.~\cite{heichinger2006genome}). Such fits
fully constrain the model parameters: fork velocity $v$, $\gamma$,
start of the S phase $t_0$, origin strengths $\lambda_i$ and
inter-origin distances $d_i$, from which we calculated
$\left<d\right>$, $\left<\lambda\right>$, $\sigma_d$ and
$\sigma_\lambda$, and simulated the duration of S phase and
replication time of each chromosome (see Appendix A and
Fig.~S8-10).
In these simulations we consider circular chromosomes with $n$
  origins, and boundary effects are tested in the Appendix B
  and Fig.~S2, and do not affect our main conclusions, indicating
  that, according to the model, the partition of the genome into 16
  unconnected chromosomes has little effect on the statistics of
  S-phase duration.
The values of $\gamma$ that were obtained as best fits of the
empirical data (Supplementary Fig.~S8) were in line with previous
analyses (e.g.~\cite{yang2010modeling,hawkins2013high}). In addition,
we found that the standard deviation of the predicted S-phase duration
decreases with the parameter $\gamma$ (Supplementary Fig.~S9), which
agrees with the finding of previous studies focused on
\emph{X.~laevis}~\cite{Bechhoefer2007,Yang2008}.

\begin{figure}
\begin{center}
  \includegraphics[width=0.5\textwidth]{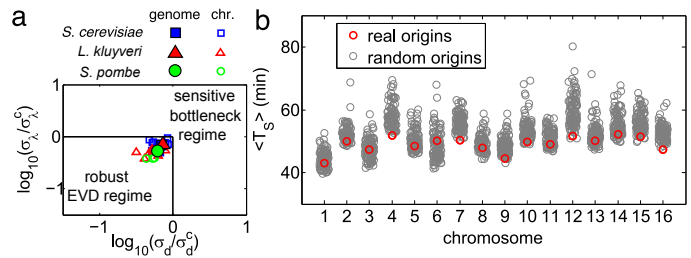}
  \end{center}
  \caption{{\bf The replication program of yeast is in the robust regime.}
    (a) Symbols are the parameters of \emph{S.~cerevisiae} (blue
    squares), \emph{L.~kluyveri} (red triangles) and \emph{S.~pombe}
    (green circles), inferred from fits with replication timing data
    from \cite{hawkins2013high}, \cite{agier2013spatiotemporal} and
    \cite{heichinger2006genome} respectively (see Supplementary Table
    S1).  Filled symbols correspond to the whole genome, hollow
    symbols to each chromosome.  (b) For each chromosome of
    \emph{S.~cerevisiae}, the average S-phase duration (y axis) is
    compared (by simulations of the model) between empirical origin
    positions and firing strengths (red circles) and randomized
    origins with empirically fixed distributions (grey circles).  }
\label{realdata}
\end{figure}

This analysis indicates that the whole-genome values of $\sigma_d$ and
$\sigma_\lambda$ measured for \emph{S.~cerevisiae}, \emph{L.~kluyveri}
and \emph{S.~pombe} place these genomes within the extreme-value
regime.  Rescaling $\sigma_d$ and $\sigma_\lambda$ by the crossover
values $\sigma^\mathrm{c}_d$ and $\sigma^\mathrm{c}_\lambda$
respectively makes it possible to compare data with different mean
$T_\mathrm{S}$. This comparison (Fig.~\ref{realdata}a) shows that not
only the genomic but also most of chromosomal parameters of
\emph{L.~kluyveri}, \emph{S.~cerevisiae} and \emph{S.~pombe} are
located in the extreme-value regime.
With the fitted parameters, most of chromosomes and genomes are
  found in the extreme-value regime (as an example, see Supplementary
  Fig.S10).
%
Interestingly, all chromosomes (and the full genome) lie close to the
transition line.  This may be a consequence of the presence of
competing optimization goals, such as replication speed (or
reliability) and resource consumption by the replication
machinery~\cite{Bechhoefer2007}.

Furthermore, we considered data of two \emph{S.~cerevisiae}
  mutants.  In one mutant, three specific origins in three different
  chromosomes (6, 7, and 10) were
  inactivated~\cite{hawkins2013high}.
  The inactivation of a specific origin slows down the replication of
  the nearby region, which might cause a bottleneck.
  Our results show that this origin mutant is still in EVD regime
  (Supplementary Fig.~S13).
  Importantly, in this case the model should be able to make a precise
  prediction for the replication profile of the chromosomes where one
  origin is inactivated.
  Supplementary Fig.~S14 shows the prediction on the replication
  profile of origin mutant strain based on the parameters fitted from
  the data of wild-type strain (except that the three inactivated
  origins are deleted from the origin list). The model prediction is
  in fairly good agreement with data. The mismatch between prediction
  and data in some regions (but not others) is an interesting feature
  revealed by the model, and may result from experimental error or
  gene-expression adaptation of the mutants~\cite{hawkins2013high}.
  The other mutant strain that we considered is \emph{isw2/nhp10},
  from the study of Vincent and coworkers~\cite{vincent2008atp}, who
  analyzed the functional roles of the Isw2 and Ino80 complexes in DNA
  replication kinetics under stress. This study compares the behavior
  of wild type (wt) strain and a \emph{isw2/nhp10} mutant in the
  presence of MMS (DNA alkylating agent methyl methanesulfonate) and
  found that S-phase in \emph{isw2/nhp10} is extended compared to the
  wt strain because the Isw2 and Ino80 complexes facilitate
  replication in late-replicating-regions and improve replication fork
  velocity. In agreement with these findings, the model fit of the
  data shows that \emph{isw2/nhp10} mutant has more inactive origins
  and smaller fork velocity. Such conditions may facilitate the onset
  of a bottleneck regime in the mutant compared to the wt strain. We
  found that~\emph{S.~cerevisiae} wt strain treated with MMS still
  falls in the extreme-value regime. Conversely, some chromosomes (e.g
  13 and 15) of the \emph{isw2/nhp10} mutant are in the bottleneck
  regime, and in this case, the whole genome (entire S-phase), is
  driven in the bottleneck regime (see Supplementary Fig.~S15).
  Strikingly, the model makes a good prediction on the replication
  profile of the \emph{isw2/nhp10} mutant, using origin firing
  strengths and the $\gamma$ values fitted from the wild-type strain
  experiments, and just adjusting two (global) parameters replication
  speed and an overall factor in all origin firing rates
  (Supplementary Fig.~S16). This provides a good cross-validation of
  the applicability of the model in a predictive framework.

A further question is whether we can detect signs of optimization in
the duration of chromosome replication.
Fig.~\ref{realdata}b compare the S-phase durations obtained from
simulations of the model in two cases: (i) by using the origin
positions and strengths from empirical data (see Supplementary
Fig.~S10), and (ii) by using a null model with randomized parameters
(both origin strengths and inter-origin distances) drawn according to
Eq.~(\ref{eq:gammas}), and preserving the empirical mean and variance.
The results show that for some of the chromosomes the average
replication timing $T_\mathrm{S}$ is close to the typical one obtained
from randomized origins (e.g., chromosomes 1,3,5,6,8,11,13 in
\emph{S.~cerevisiae}).  For other chromosomes (e.g., 2,4,7,10,12,15,16 in
\emph{S.~cerevisiae}) the empirical average $T_\mathrm{S}$ is instead
very close to the minimum reachable within their ensemble of
randomizations.
Remarkably, chromosomes with higher average replication timing in the
randomized ensemble seem to be more subject to pressure towards
decreasing their average $T_\mathrm{S}$ (Supplementary Fig.~S11).
This result suggests that the whole replication program may be under
selective pressure for fast replication.

\section{DISCUSSION}

The core of our results are analytical estimates that capture the
cell-to-cell variability in S-phase duration based on the measurable
parameters of replication kinetics.
%
{Extreme-value statistics has been applied to DNA replication
  before~\cite{Yang2008,Bechhoefer2007}, but only to the case of
  organisms like \emph{X.~laevis}, where origin positions are not
  fixed and there is no spatial variability of initiation rates. To
  our knowledge, this method has not been applied systematically to
  fixed-origin organisms such as yeast.
%
%
More specifically ref.~\cite{Yang2008} explores the case of a perfect
lattice of equally spaced discrete origins with fixed and equal firing rates,
but does not address the role of the variability of inter-origin
replication times due to randomness in firing rates and inter-origin
distance, which is relevant for fixed-origin organisms.
Another difference is that the authors of
ref.~\cite{Yang2008,Bechhoefer2007} derive the coalescence
distribution starting from their model, while here we assume a
stretched-exponential, motivated by data analysis.
Since their distribution is more complex (although the model is
simpler), EVD estimate leads to a formula linking the parameters of
the Gumbel distribution to the initiation parameters in the form of an
implicit equation, that needs to be solved numerically.
Conversely, the assumption that the shape of the distribution of $T_i$
is given (and estimated from data), gives an explicit relationship
between the parameters describing the $T_i$ distribution and the
Gumbel parameters, leading to simpler formulas and applicability to
the case of discrete origins with different spacings and firing
rates. The parameters of the $T_i$ distribution have then to be
related to the microscopic parameters (See Appendix D).
}

It is important to note that an approach based on extreme-value
distribution theory is general~\cite{Bechhoefer2007}.  Simulations
(including the model used here) are based on specific assumptions that
are often not simple to test and many models on the market use
slightly different assumptions.
{
Instead, the extreme-value estimates are robust to different shades of
assumptions used in the models available in the literature, and thus
more comprehensive.
}
Our estimates reveal universal behavior in the distribution of S-phase
duration.  There is a prescribed relation between mean and
variance of S-phase duration, defining a ``scaling'' behavior for its
distribution. Such universality has been observed in cell-cycle
periods and cell size~\cite{Kennard2016,Giometto2013}.
%
{
Qualitatively, we expect the same universality to hold in a regime
when origins have less than 100\% efficiencies, and some may not fire
at all during S-phase.  Origins that fire only in a fraction of the
realizations are accounted for in our simulations, but they entail
second-neighbour effects that are not currently accounted in our
estimates.
}

There are hundreds of origins in a genome, but our
analysis shows that the relevant parameters to capture the overall
behavior are the means and variances of inter-origin distances and
origin firing rates.
Specifically, we find that two regimes describe most of the
phenomenology, and they depend on the values of these effective
variables.
{Importantly, the regimes identified here differ from those
  identified in ref.~\cite{Yang2008}, which just identify a critical
  spacing between discrete (equally spaced) origins, for which
  replication timing starts to be linear with inter-origin distance.
}

 {The notion that the last regions to replicate may tend to be
  different in every cell (our ``extreme-value'' regime) has been
  proposed already by Hawkins and
  coworkers~\cite{hawkins2013high}. The opposite regime where some
  specific regions tend to always replicate last ('bottleneck
  region'), has been proposed for mammalian common fragile
  sites~\cite{Letessier2011}. Such regions of slow replication,
  pausing and frequent termination have also been described in
  yeast~\cite{Cha2002,Ivessa2003,Fachinetti2010,hawkins2013high}.
  These studies make it plausible to think that both extreme-value and
  bottleneck regimes may apply to yeast, despite our analysis based on
  replication kinetics data indicating some pressure towards the
  extreme-value regime.
  Another important case for what concerns replication termination is
  the rDNA locus, which cannot be analyzed in replication kinetics data
  based on microarrays / sequencing data due to its repetitive nature
  ($\thicksim$150 identical copies in yeast). However, the large inter-origin
  distances, pseudo-unidirectional replication and epigenetic control
  of origin firing in this locus~\cite{Pasero2002} make it a good
  candidate for the last sequence to replicate in yeast. }

Importantly the model used here is similar to a set of previous
  studies, which have tested this approach and validated it with
  experimental data
  ~\cite{Yang2008,yang2010modeling,bechhoefer2012replication,retkute2011dynamics,retkute2012mathematical,hawkins2013high}. Our
  analysis of S-phase duration in single cells is generic, and
  expected to be robust to variations model details.
  The mutant data sets analyzed here also support the predictive power
  of the model in presence of perturbations and parameter changes, and
  hence validate the use of the model in a predictive framework.
Our predictions are compatible with the available values for
  average S-phase duration, which can be roughly estimated through
flow cytometry~\cite{hawkins2013high,agier2013spatiotemporal}, and
corresponds well to the values obtained by the model (around 60
minutes for~\emph{S.~cerevisiae}). Other yeast studies found smaller
values in other conditions~\cite{Magiera2014d}, which would be
interesting to study with the model.
Additionally, we provide a prediction for the cell-to-cell variability
of S-phase duration, which is an important step of the cell cycle.
Indeed, completion of replication needs to be coordinated with growth
and progression of the cell cycle
stages~\cite{Schmoller2015a,Skotheim2013}. Cell-to-cell variability in
replication kinetics makes the S phase subject to inherent
stochasticity.
Experimentally, measuring the cell-to-cell variation of the S-phase duration
is a challenge.
While some studies exist using mammalian (cancer) cell
  lines~\cite{Hahn2009}, they currently do not have the precision
needed to allow a quantitative match with models.
However, we expect that such measurements will become available in the
near future, thanks to rapidly developing methods of single-cell
biology{~\cite{Bajar2016}}.
Our predictions define some key properties of the replication period
that may be tested with, e.g., single-cell studies in budding yeast,
using the parameters available from replication kinetics studies.  In
this model the S phase is (by itself) a ``timer'', so its connection
to cell size homeostasis must be affected by external
mechanisms~\cite{Schmoller2015a}. S-phase duration has been measured
on single \emph{E.~coli} cells, and found to be unlinked to cell
size~\cite{Adiciptaningrum2015}.
%
Interestingly, our predictions of S-phase duration and
  variability as a function of chromosome copy numbers (Supplementary
  Fig.~S12) might apply to cancer cell lines with different levels of
  aneuploidy~\cite{Hahn2009}.
Finally, there is the possibility of applying this framework to
describe relevant perturbations~\cite{Koren2010,Gispan2014}. This
could also help elucidate how response to DNA damage affects the
replication timing and its variability across cells.

Intriguingly, we also found evidence of bias towards faster
replication in empirical chromosomes compared to randomized ones.
Thus, our overall findings support the hypothesis of a possible
  selective pressure for faster replication, and against bottlenecks.
Other approaches have assumed optimization for faster replication and
looked for optimal origin placement~\cite{karschau2012optimal} or
found other signs of optimality in similar
data~\cite{yang2010modeling}. Our results are in line with these
findings, and isolate a complementary direction for such optimization.
All these considerations support the biological importance of
replication timing of inter-origin regions and its variability.
However, the sources of the constraints remain an open
question. Clearly, overall replication speed can increase indefinitely
by increasing origin number and initiation rates. However, there are
likely yet-to-be-characterized tradeoffs in these quantities, that
prevent this from happening, and force the system to optimize the
duration of replication in a smaller space of parameters. The
molecular basis for such constraints likely lies at least in part in
the finite resources available for initiation
complexes~\cite{Das2015}.

\begin{acknowledgments}
We are grateful to Gilles Fischer, Nicolas Agier, Alessandra Carbone
and Renaud Dessalles for useful discussions.
QZ was supperted by the LabEx CALSIMLAB, public grant
ANR-11-LABX-0037-01 constituting a part of the ``Investissements
d'Avenir'' program (reference : ANR-11-IDEX-0004-02; YK).
\end{acknowledgments}

\appendix

\section{Fitting replication timing data from experiments using the
  model}

This section describes our fitting procedure based on the model. The
fitted parameters were used in simulations of genome replication
kinetics can giving the distribution of S-phase duration and of
replication time of one chromosome (Fig.~\ref{fit2}).

We used flow cytometry (FACS) data to re-normalize replication timing
as follows.  If the base line value of average DNA copy-number $a$ is
remarkably larger than 1, and/or its plateau value $b$ is remarkably smaller
than 2, we use the formula
$y=a+\frac{(b-a)(t-T_0))^r}{(t-T_0)^r+(t_c-T_0)^r}\theta(t-T_0)$ to
fit the FACS data and normalize replication timing data by $\phi_{\rm
  norm}(x,t)=1+\frac{\phi(x,t)-a}{b-a}$, where $\phi$ is the
replication probability function~\cite{agier2013spatiotemporal}.

We used fixed origin locations from the literature and optimized the
fit for the parameters $\gamma$, $T_0$, $v$ and $\lambda_i$
iteratively. The objective function was defined as the L2 distance
(the average of squared differences) of the experimental and
theoretical replication probability timing profile (Fig.~\ref{fit2}),
i.e., as $\sqrt{\sum_i\sum_j(\phi_{model}(x_i,t_j)-\phi_{exp.}(x_i,t_j))^2/(N_xN_t)}$,
where $N_x$ and $N_t$ are the numbers of the measured loci and time
points respectively.

Initialization of the parameters for the fits was performed as
follows. Firing rate exponent $\gamma$ and fork velocity $v$ were
initialized at arbitrary values (typically $\gamma$ at 0, $v$ at 2
kb/min). The start of S phase $T_0$ was initially set when genome copy
number from the normalized FACs data (from the interval $[a,b]$ to
$[1,2]$) is first larger than a fixed threshold (e.g. 1.05) and each origin
strength $\lambda_i$ starts from the value fitted with the time-course
data at this origin.

Fitting was performed with following iterative rule. 1) for a
parameter x, assume it has a step length $\Delta_x$, and a memorized
step length $\Delta_x^{'}=2\Delta_x$, 2) set $r=\Delta_x/\Delta_x^{'}$
and $\Delta_x^{'}=\Delta_x$, if $x+\Delta_x$ gives a better fit than
$x$, let $x=x+\Delta_x$, otherwise (i) if $|r|=1$, we update
$\Delta_x \rightarrow \Delta_x/2$ (ii) if $|r|=0.5$, set $\Delta_x
\rightarrow -\Delta_x$; 3)
repeat 2) until the termination condition is satisfied.  $\lambda_1$,
$\lambda_2$, ..., $\lambda_n$ for each chromosome are updated
iteratively given $\gamma$, $v$ and $T_0$ and in each iteration, one
$\lambda_i$ is chosen randomly to be updated. $T_0$ is updated
iteratively given $\gamma$ and $v$.  $v$ is updated iteratively given
$\gamma$. For $\gamma$, we tested some discrete values between 0 and 3.
Supplementary Fig.~\ref{fit1}a,b indicate the best fit value of $\gamma$
for \emph{S.cerevisiae} and \emph{L.kluiveri},
and Supplementary Fig.~\ref{fit1}c shows one example of the best fit.

\section{Role of chromosome boundaries in replication timing}

In some simulations, we used circularized chromosomes for easier
comparison with the analytical estimates, but relative to a circular
chromosome, a linear chromosome has lower symmetry because of the
boundary at both ends.  To verify that this assumption does not
qualitatively affect the results, we circularized the empirical
\emph{S.cerevisiae} chromosomes by linking their ends respectively,
and simulated their replication kinetics with the estimated
parameters. The results (Fig.~\ref{boundaryEffect}) show that the
circularized chromosomes always replicate faster than the linear
chromosomes, but their durations do not differ much (the average
deviation is in all cases less than 15\%).


\section{Determination of the parameters $\alpha$, $\beta$ and $t_0$
  in the formula for the distribution of $T_i$}

Eq.~3 in the main text, describing the replication timing of one
inter-origin region contains the parameters $\alpha$, $\beta$ and
$t_0$, which need to be related to the biologically measurable
parameters (inter-origin distance and origin rates).  To estimate such
parameters for the distribution of $T_i$ we used two methods. The
first is a fit of all the $T_i$ data taken from the simulation of the
given chromosome, and the second is to fit the specific
$T_i$ data (replication times of the central inter-origin region) extracted
 from simulation of a linear chromosomal fragment where inter-origin distances and origin
strengths are sampled from known distributions (different samples for different runs of the simulation).
In this second
method, each run of the simulation is carried out considering
inter-origin distances and origin strengths with the same averages as
the original chromosome.
Both methods give the same distribution for $T_i$, which agrees
very well with Eq.~3 of the main text (See Fig.~\ref{Assumption_Ti}).

We mainly used the second method since it does not depend on origin
configuration of the original chromosome. The detailed procedure is
the following. First, we defined a characteristic distance
$d_c=(\frac{\gamma+1}{\left<\lambda\right>}\log(\frac{1}{1-x}))^{\frac{1}{1+\gamma}}v$,
where $x<1$ (e.g. 0.99) and assume $n_c=\min(\lfloor
d_c/\left<d\right>\rfloor+1,\lfloor n/2 \rfloor)$+1. Then we produced
a linear chromosomal fragment with $2n_c$ origins, in which two origins are
always located at the ends. Next, we simulated many realizations for
the replication of this chromosome. In each simulation run, we sampled
inter-origin distance $d_i$, origin strength $\lambda_j$ and origin
firing time $t_f^{(j)}$ from
$\Gamma(\frac{\left<d\right>^2}{\sigma^2(d)},\frac{\left<d\right>}{\sigma^2(d)})$,
$\Gamma(\frac{\left<\lambda\right>^2}{\sigma^2(\lambda)},\frac{\left<\lambda\right>}{\sigma^2(\lambda)})$
and $f(t)=\lambda_i t^\gamma\theta(t){\rm
  exp}(-\lambda_i\frac{t^{\gamma+1}}{\gamma+1})$ respectively, where
$i\in\{1,2,...,2n_c-1\}$ and $j\in\{1,2,...,2n_c\}$. The statistics
over different realizations gives the distribution of the replication
time of the central inter-origin region ($T_{n_c}$), which was fitted
with Eq.~\ref{S_ti} to obtain $\alpha$, $\beta$ and $t_0$.

\makeatletter
\renewcommand{\theequation}{S\@arabic\c@equation}
\makeatletter

\section{Analytical derivation of an approximate distribution of S-phase
  duration $T_S$ based on extreme value theory.}

This section gives further details on the analytical calculation for
the extreme-value estimate of the distribution of S-phase duration.
We assume that replication timing of one inter-origin region $T_i$
obeys the stretched exponential distribution
\begin{equation}
  F(t)=P(T_i<t)=1-e^{-\alpha (t-t_0)^{\beta}} \ ,
  \label{S_ti}
\end{equation}
where $t\geqslant t_0$ and $\alpha>0$. The parameters $\alpha$,
$\beta$ and $t_0$ were obtained as described in the previous
section. We define ${M_n}=\max\left(T_1, T_2,...,T_n\right)$. By
taking $ a_n=1/(\alpha^{1/\beta}\beta(\log n)^{1-1/\beta})$ and
$b_n=(\log n/\alpha)^{1/\beta}+t_0$, and applying the
Fisher-Tippett-Gnedenko theorem, we can prove that
\begin{equation}
\lim\limits_{n\rightarrow\infty}P((M_n-b_n)/a_n\leq
t)=\exp(-\exp(-t))\triangleq G(t) \ ,
\end{equation}
where $G(t)$ is the standard Gumbel distribution.

When $n$ is sufficiently large, we can make the approximation
$P((M_n-b_n)/a_n\leq t)\approx G(t)$.  If we define $\tilde{t}=a_n
t+b_n$, we have $P(M_n\leq \tilde{t})\approx G((\tilde{t}-b_n)/a_n)$.

Finally, we can represent the distribution of $T_S$ (=$M_n$)
approximately as
\begin{equation}
\begin{aligned}
 & P(T_S\leq t)  \approx  \exp(-\exp(-\frac{t-b_n}{a_n})) \\
  &= \exp\left\{-\exp\left[\beta\log
  n\left(1-(\alpha/\log
  n)^{1/\beta}(t-t_0)\right)\right]\right\} \\
  \end{aligned}
\end{equation}
%
{ Here $n$ is the origin number, and $\alpha$, $\beta$ and $t_0$
  are connected to the model parameters describing replication
  kinetics, $v$, $\gamma$, inter-origin distances ($d_1, d_2, ...,
  d_n$) and origin strengths ($\lambda_1,\lambda_2,...,\lambda_n$).}

We now discuss how $\alpha$, $\beta$ and $t_0$ can be expressed
  as functions of simplified parameters by numerically solving some
  approximate equations.  We consider a ``characteristic''
  inter-origin region with the distance $\left<d\right>$ and origin
  strength $\left<\lambda\right>$, and we assume that the replication
  of the inter-origin region is mainly carried out by the forks
  originated from the two nearest origins, both of which are typically
  activated, Thus we have
\begin{equation}
  T_i\approx \left<d\right>/2v+(t_f^{l}+t_f^{r})/2, \label{Ti_appro}
\end{equation}
where $t_f^l$ and $t_f^r$ are the firing time of the left origin and
the right origin respectively. Since $t_0$ is the minimal replication time
of inter-origin region and the firing time has zero as a lower bound,
one has
\begin{equation}
  t_0=\min(T_i)=\left<d\right>/2v.  \label{t0}
\end{equation}
From equation \ref{Ti_appro}, we can further obtain
\begin{equation}
\left<T_i\right> \approx \left<d\right>/2v+\left<t_f\right>  \label{eqs1}
\end{equation}
and
\begin{equation}
\sigma(T_i)\approx \sigma(t_f) \label{Ti_sd}
\end{equation}
In addition, we have
\begin{equation}
\left<T_i\right>=\alpha^{-\frac{1}{\beta}}\Gamma\left(\frac{1}{\beta}+1\right)+t_0,
\end{equation}
\begin{equation}
  \sigma(T_i) =
  \alpha^{-\frac{2}{\beta}}
  \left[\Gamma\left(\frac{2}{\beta}+1\right) -
  \Gamma^2\left(\frac{1}{\beta}+1\right)\right],
\end{equation}
\begin{equation}
  \left<t_f\right> =
  \left(\frac{\gamma+1}{\left<\lambda\right>}\right)^{\frac{1}{\gamma+1}}\Gamma\left(\frac{\gamma+2}{\gamma+1}\right),
\end{equation}
and
\begin{equation}
\sigma(t_f)=\left(\frac{\gamma+1}{\left<\lambda\right>}\right)^{\frac{1}{\gamma+1}}\sqrt{
\Gamma\left(\frac{\gamma+3}{\gamma+1}\right)-\Gamma^2\left(\frac{\gamma+2}{\gamma+1}\right)}
\label{eqs2}\end{equation}
Based on equations \ref{t0}-\ref{eqs2}, $\alpha$ and $\beta$ can be
numerically solved as functions of $v$, $\gamma$, $\left<d\right>$ and
$\left<\lambda\right>$.  Our simulations in the EVD regime, and using
empirically realistic values of the parameters are in line with
equations \ref{t0}-\ref{Ti_sd}.

\nocite{*}

\bibliography{S-phase}

\clearpage

\makeatletter \renewcommand{\thefigure}{S\@arabic\c@figure}
\setcounter{figure}{0}
\makeatletter

\onecolumngrid
\section*{Supplementary Figures and Tables}

\label{testgamma}
\begin{figure}[h!]\centering
  \includegraphics[width=1\textwidth]{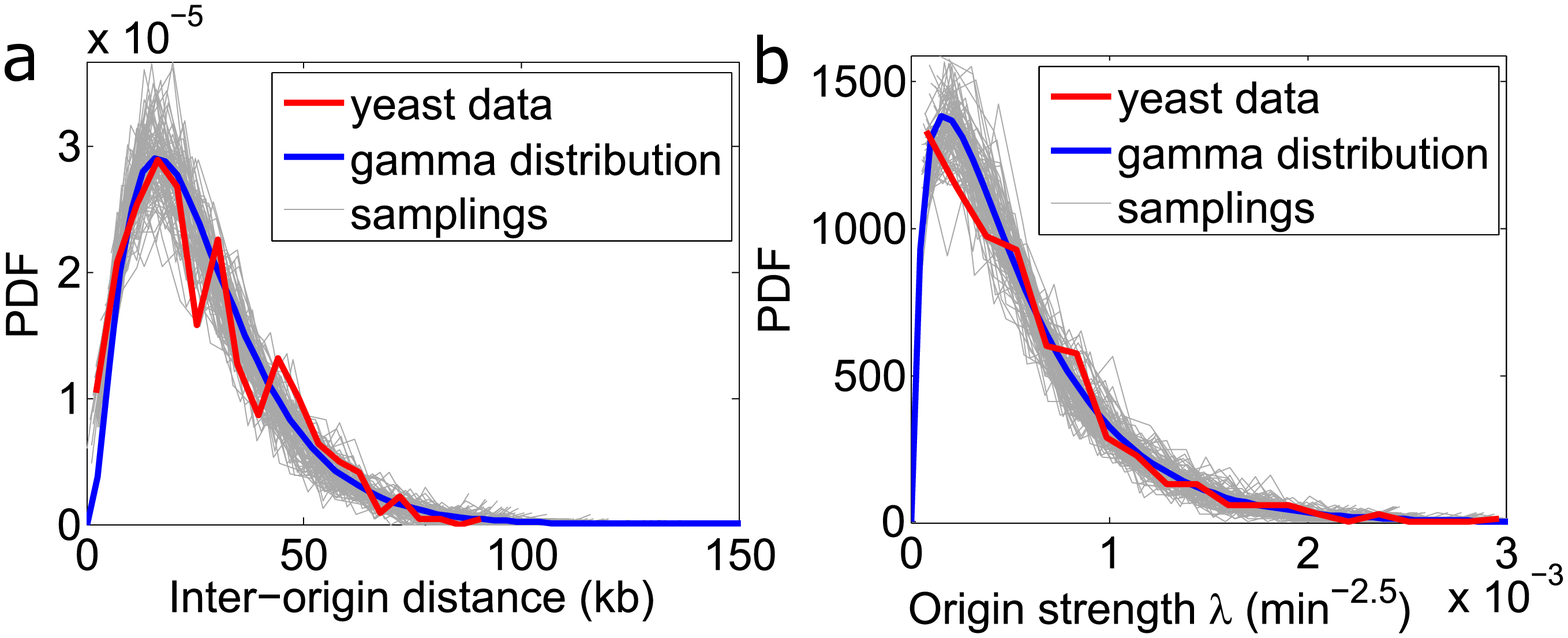}
  \caption{\textbf{The hypothesis of gamma-distributed inter-origin
      distances and origin firing rates used to generate randomized
      chromosomes is in line with empirical data}. The plots compare
    inter-origin distances (a) and firing rates (b) distributions used
    for the model (blue continuous line) with \emph{S. cerevisiae}
    data from ref.~\cite{hawkins2013high} (red line), and 100
    samplings of the assumed distributions with the same number of
    instances as the empirical case (thin grey lines).  Empirical
    firing rates were inferred setting $\gamma=1.5$ (the best-fit
    value for the data in
    ref.~\cite{hawkins2013high}.} \label{testgamma}
\end{figure}

\clearpage

\label{boundaryEffect}
\begin{figure}[h!]\centering
  \includegraphics[width=0.5\textwidth]{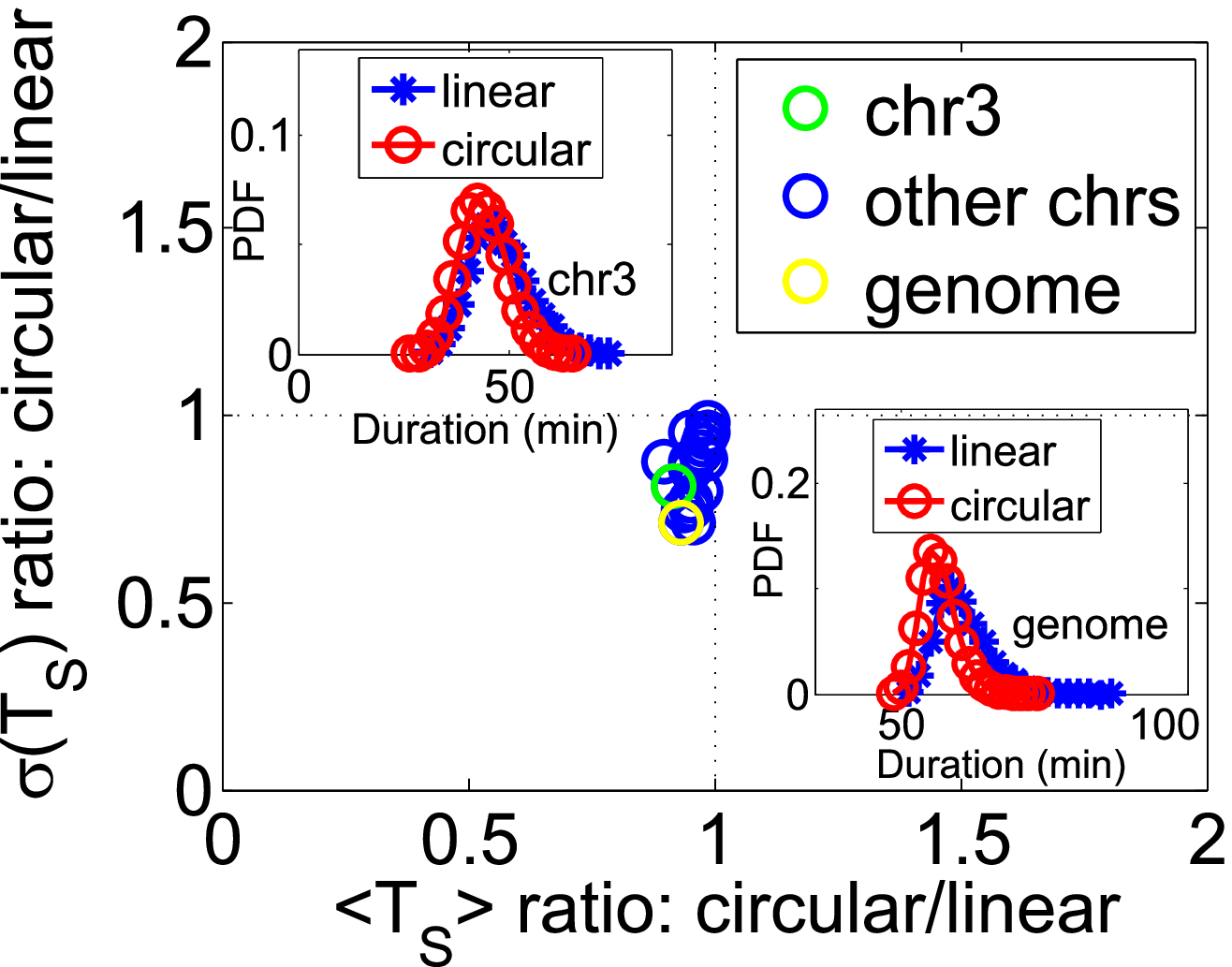}
  \caption{\textbf{Comparison of S-phase duration of
      \emph{S.cerevisiae} chromosomes and genome and their circularized versions
      indicates that the boundary effect on replication timing is
      small. } Circular chromosomes were obtained
    by linking two ends of the linear chromosome. The circular genome was gotten
    by linking all the linear chromosomes via their ends successively.
    Ratio of $T_S$ average (SD) between \emph{S.cerevisiae}
    linear chromosomes and the genome and the circularized versions is close to 1.  The insets show
    that the distribution of $T_S$ of chromosome 3 and the genome and their
    circularized versions are similar. The parameters giving best fit
    to \emph{S. cerevisiae} data from ref.~\cite{hawkins2013high} were
    used (in particular, $\gamma=1.5$). }. \label{boundaryEffect}
\end{figure}

\clearpage

%

\label{Assumption_Ti}
\begin{figure}[h!]\centering
  \includegraphics[width=1\textwidth]{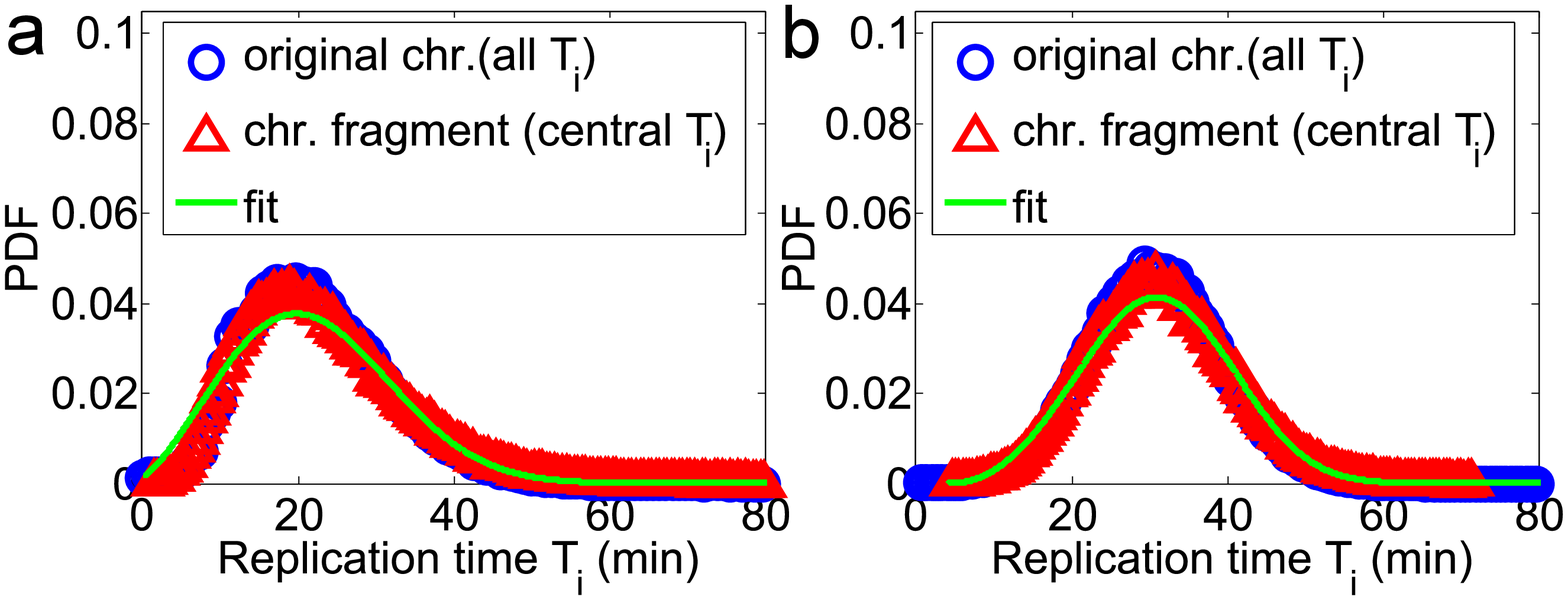}
  \caption{\textbf{Justification of the assumption for the
      inter-origin replication timing distribution (Eq.~3 of the main
      text).} We used two methods of obtaining the data for the
    distribution of replication time ($T_i$) of inter-origin regions,
    both of which are in good agreement with the theoretical
    formula. Blue circles: distribution obtained by the simulation of
    an circular chromosome (original chromosome) where origin
    strengths and inter-origin distances are sampled with Eq.2; red
    triangles: distribution of replication time of the central
    inter-origin region in a linear chromosomal fragment where origin
    strengths and inter-origin distances are sampled with Eq.2 in each
    run of the simulation; the continuous line is a fit with
    Eq.~3. For (a), chromosome parameters: $\gamma=0$, $n=20$
    (original) or $16$ (linear fragment), $v=1.88\ kb/\min$,
    $\left<d\right>=28.13\ kb$, $\sigma(d)=13.46\ kb$,
    $\left<\lambda\right>=0.045\ \min^{-1}$, $\sigma(\lambda)=0.036\
    \min^{-1}$, and the fitted parameters: $\alpha=3.72\times 10^{-4}\
    \min^{-\beta}$, $\beta=2.42$, $t_0=-1.07\ \min$. For (b),
    chromosome parameters: $\gamma=1.5$ (best fit), $n=20$ (original)
    or $10$ (linear fragment), $v= 1.81\ kb/\min$, ,
    $\left<d\right>=28.13\ kb$, $\sigma(d)=13.46\ kb$,
    $\left<\lambda\right>=6.17\times 10^{-4}\ \min^{-2.5}$,
    $\sigma(\lambda)= 5.53\times 10^{-4}\ \min^{-2.5}$, and the fitted
    parameters: $\alpha=1.79\times 10^{-5}\ \min^{-\beta}$,
    $\beta=3.21$, $t_0= 4.16\ \min$.  } \label{Assumption_Ti}
\end{figure}

\clearpage
\label{Assumption_Ts}
\begin{figure}[h!]\centering
  \includegraphics[width=1\textwidth]{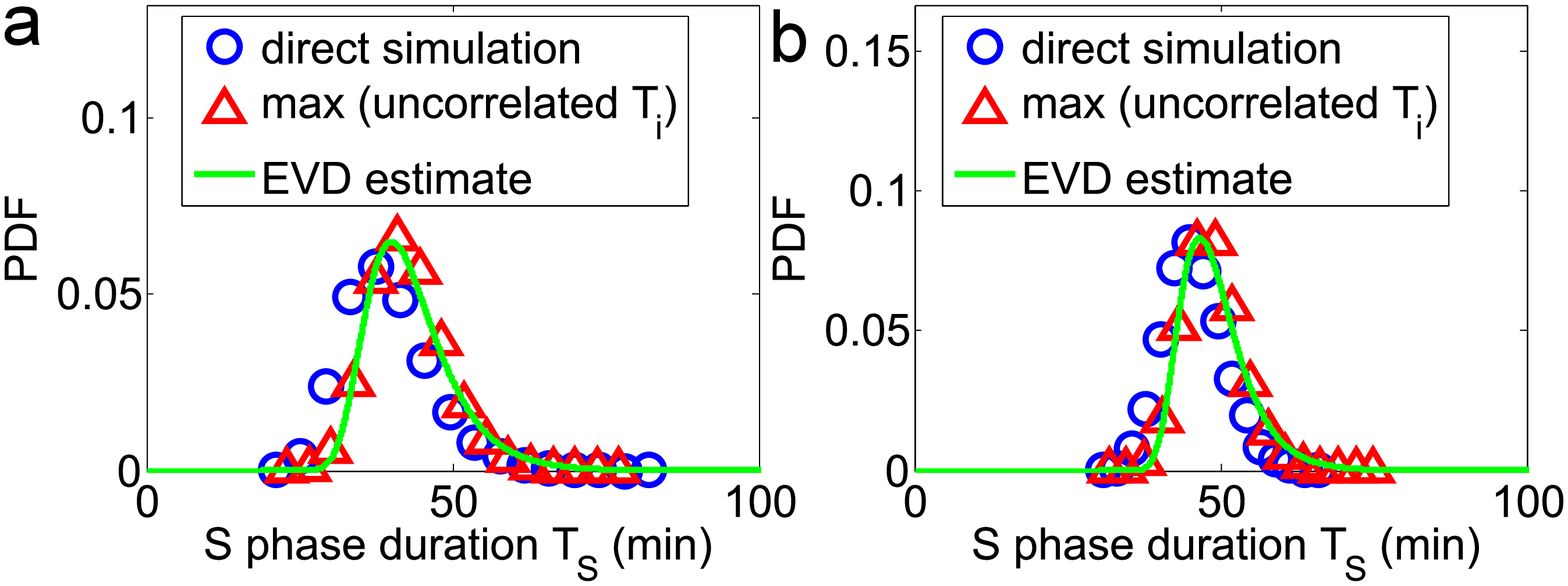}
  \caption{\textbf{The correlation between replication times of
      adjacent inter-origin regions has little effect on the
      distribution of S-phase duration $T_S$.} The plot shows the
    distribution of S-phase duration from direct simulation (blue
    circles; correlated $T_i$) compared to sampling of
    $T_i(i\in\{1,2,...,n\})$ from $F(t)=1-e^{-\alpha (t-t_0)^{\beta}}$
    independently and taking their maximum (red triangles;
    uncorrelated $T_i$). Both methods agree well with the EVD estimate
    based on Eq.~4 of the main text (green continuous line). The plots
    refers to a circular chromosome with two different parameter sets,
    compatible with yeast data: (a) $\gamma=0$, $n=20$, $v=1.88\
    kb/\min$, $\left<d\right>=28.13\ kb$, $\sigma(d)=13.46\ kb$,
    $\left<\lambda\right>=0.045\ \min^{-1}$, $\sigma(\lambda)=0.036\
    \min^{-1}$, (b) $\gamma=1.5$, $n=20$, $v= 1.81\ kb/\min$,
    $\left<d\right>=28.13\ kb$, $\sigma(d)=13.46\ kb$,
    $\left<\lambda\right>=6.17\times 10^{-4}\ \min^{-2.5}$,
    $\sigma(\lambda)= 5.53\times 10^{-4}\ \min^{-2.5}$. Origin
    strengths and inter-origin distances are sampled with Eq.~2 of the
    main text.  }\label{Assumption_Ts}
\end{figure}

\label{twoRegime}
\begin{figure}[h!]\centering
 \includegraphics[width=1\textwidth]{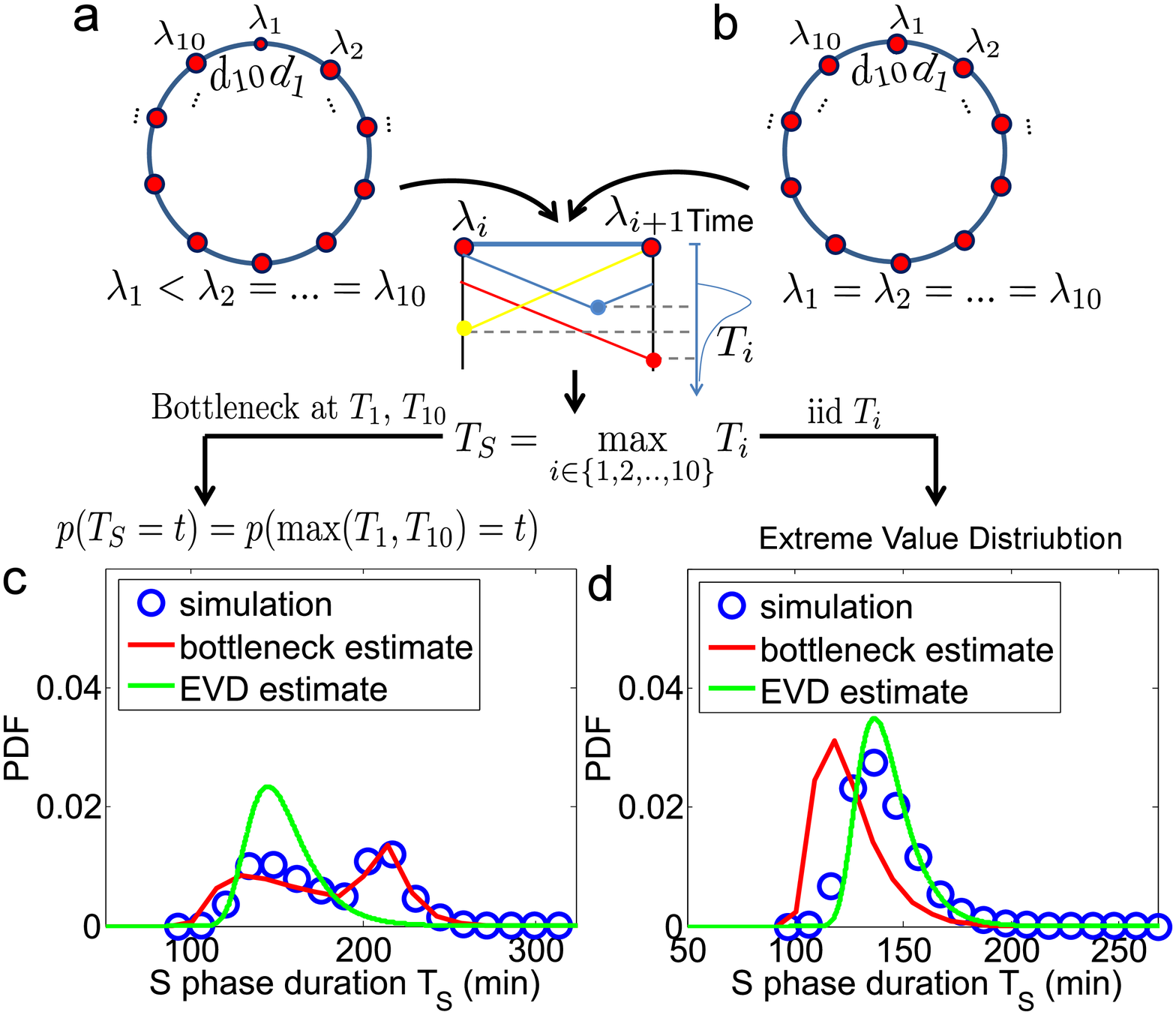}
 \caption{{\bf Replication regimes determined by firing rates.}  (a)
   Due to a single slow-firing origin, the two neighboring bottleneck
   inter-origin regions (labelled by the index 10 and 1 in panels a
   and b) typically complete replication much later than the
   rest. Hence, $T_\mathrm{S}$ will be typically equal to
   $\max(T_1,T_{10})$ (origin strengths in the example are
   $\lambda_i=0.055\ \min^{-1}$ for all origins except
   $\lambda_1=0.0055\ \min^{-1}$).  (b) If the replication times of
   all inter-origin regions are comparable, and they are considered
   independent and identically-distributed (iid) random variables, the
   distribution of $T_\mathrm{S}$ can be obtained by
   extreme-value-distribution (EVD) theory (origin strengths are
   $\lambda_i=0.05\ \min^{-1}$).  Simulations of the model (blue
   circles), when one inter-origin distance is much larger than the
   others (c), and when all inter-origin distances and strengths are
   comparable (d), agree with the corresponding analytical
   calculations (red and green curves).  (Origin number $n=10$
   origins, fork velocity $v=1\ kb/\min$, origin strength $d_i=200\
   kb$.)  }
	\label{twoRegimes}
\end{figure}

\clearpage

\label{Ti_dis_eg}
\begin{figure}[h!]\centering
 \includegraphics[width=1\textwidth]{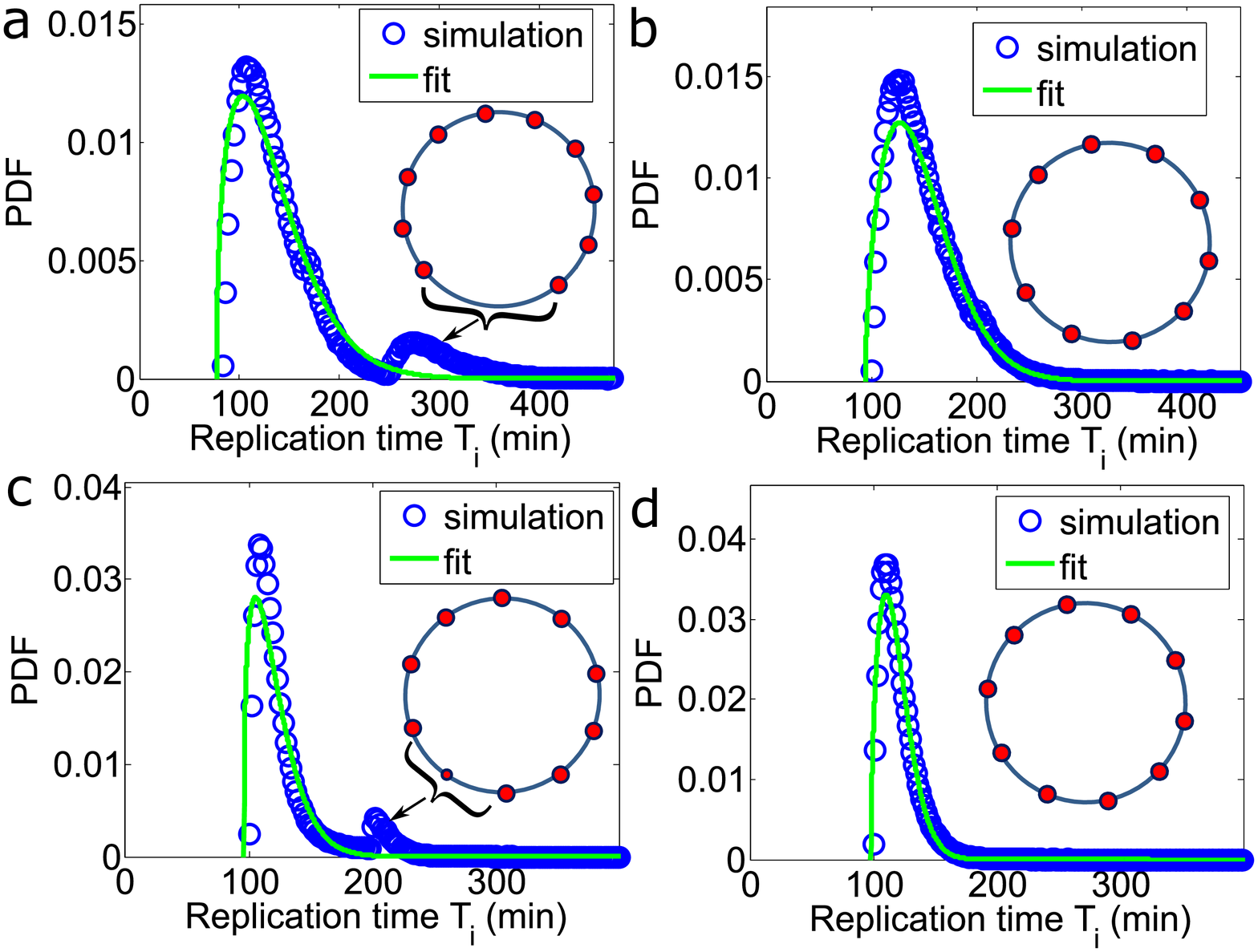}\\
 \caption{\textbf{In the bottleneck regime, the slowest region in
     replication causes the appearance of small peaks in the right
     tail of the distribution of $T_i$, leading to the failure of the
     EVD estimate.} The plots come from simulations with parameter
   sets shown in Fig.~S5 and in Fig.~2 of the main text.  For the
   bottleneck cases shown in Fig.~2 of the main text (a) and Figure S5
   (c), a small peak emerges in the right tail of $T_i$ distribution
   due to the slowest replication of the bottleneck regions.
   Conversely, in the EVD regime, the right peak does not exist (b,d).
 }\label{Ti_dis_eg}
\end{figure}

\clearpage

\label{perturbation}
\begin{figure}[h!]\centering
 \includegraphics[width=1\textwidth]{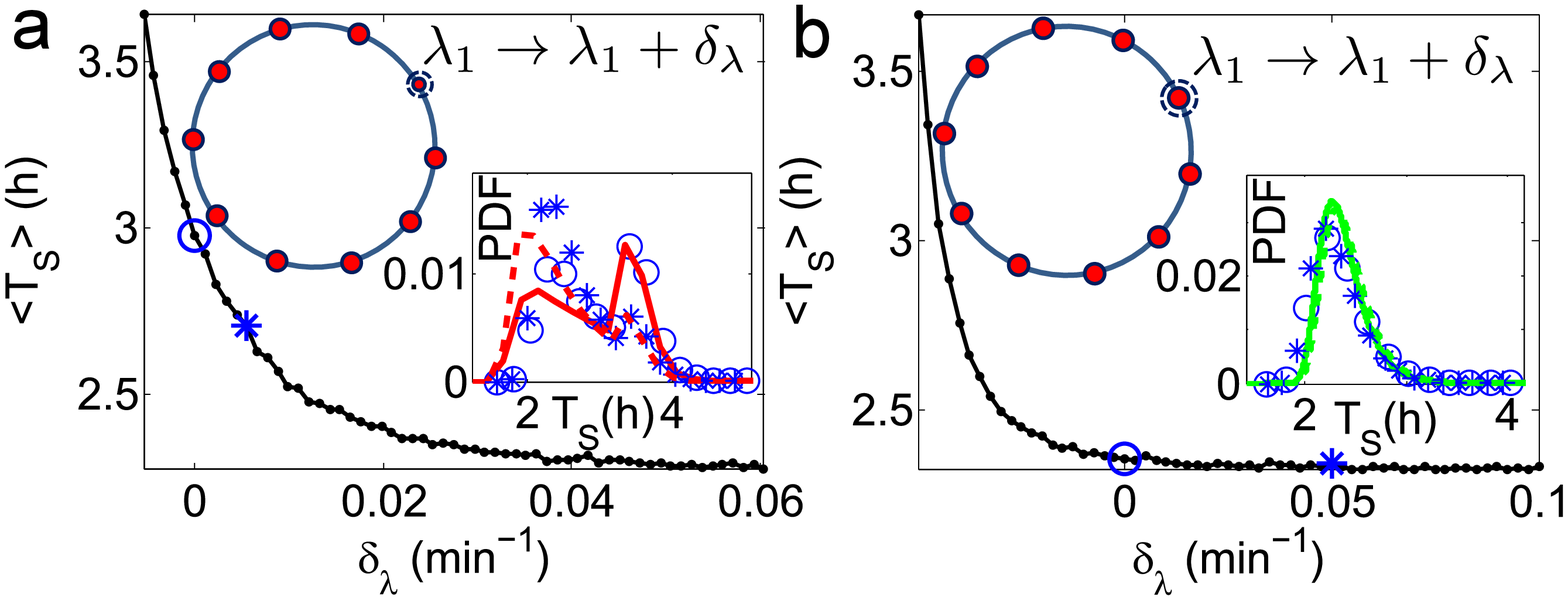}
 \caption{{\bf Effects of perturbations of a pair of inter-origin
     regions on S-phase duration.}  (a) The bottleneck regions of the
   chromosome shown in Fig.~\ref{twoRegimes}a are perturbed by
   increasing the strength of origin 1 by $\delta_\lambda$ (i.e.,
   $\lambda_1\rightarrow \lambda_1+\delta_\lambda$). The inset shows
   that the perturbation changes the distribution of $T_\mathrm{S}$
   (circles are simulations for the unperturbed chromosome, and
   stars correspond to $\delta_\lambda=\lambda_1$; the two curves
   are the analytical estimates in the bottleneck regime).  (b) The
   same perturbation as in (a) is performed on the strength of one
   origin of the chromosome shown in Fig.~\ref{twoRegimes}b, which
   lies in the EVD regime.  Symbols are as in (a).  The distribution
   of $T_\mathrm{S}$ is robust to this perturbation.  }
\label{perturbation}
\end{figure}

\clearpage

\label{fit1}
\begin{figure}[h!]\centering
 \includegraphics[width=1\textwidth]{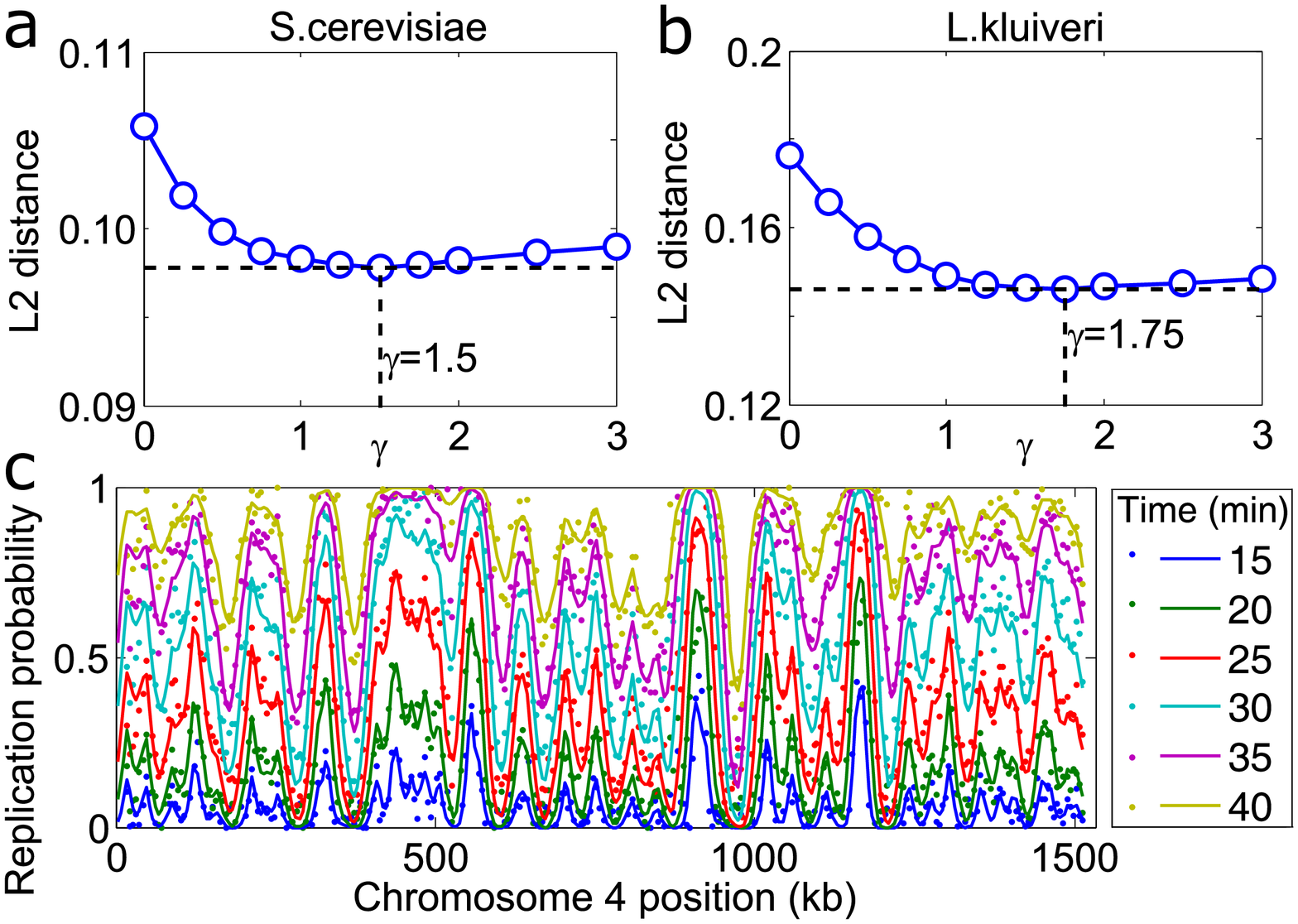}
 \caption{\textbf{The goodness of the fit of the model with the
     empirical data depends on exponent factor $\gamma$ and with the
     best $\gamma$, the model can be efficiently fit to the empirical
     replication data.} The empirical data of \emph{L.kluveri} and
   \emph{S.cerevisiae} are from
   ref.~\cite{hawkins2013high,agier2013spatiotemporal}.  (a,b) The L2
   distance between theoretical and empirical replication probability
   profiles is minimized at $\gamma=1.5$ (for \emph{S.cerevisiae}) or
   $\gamma=1.75$ (for \emph{L.kluveri}) (c) The model gives a good fit
   to the empirical replication probability $\phi(x,t)$ from
   \emph{S. cerevisiae} chromosome 4. Dots and continuous lines
   indicate experimental and theoretical data respectively, which are
   both averaged with bins of 5kb. Different colors indicate different
   measurement times.}\label{fit1}
\end{figure}

\clearpage

\label{sigma_Ts_gamma}
\begin{figure}[h!]\centering
 \includegraphics[width=0.5\textwidth]{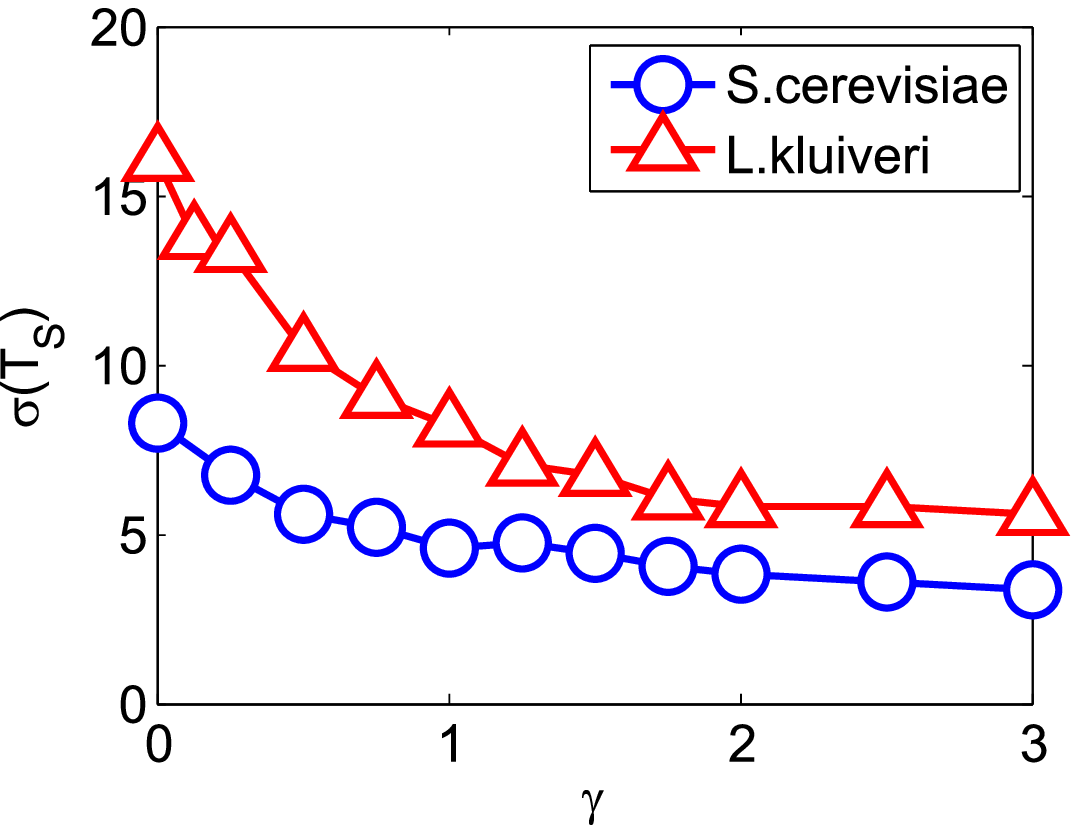}
 \caption{\textbf{The standard deviation of S-phase duration $T_S$ of
     \emph{S. cerevisiae} and \emph{L.kluveri} decreases with the
     parameter $\gamma$.} The plot is obtained from simulations with
   the best-fitting parameters of empirical data, using data from
   ref.~\cite{agier2013spatiotemporal,hawkins2013high} (See
   Fig.~\ref{fit2} and \ref{fit1})}\label{sigma_Ts_gamma}
\end{figure}

\clearpage

\label{fit2}
\begin{figure}[h!]\centering
 \includegraphics[width=1\textwidth]{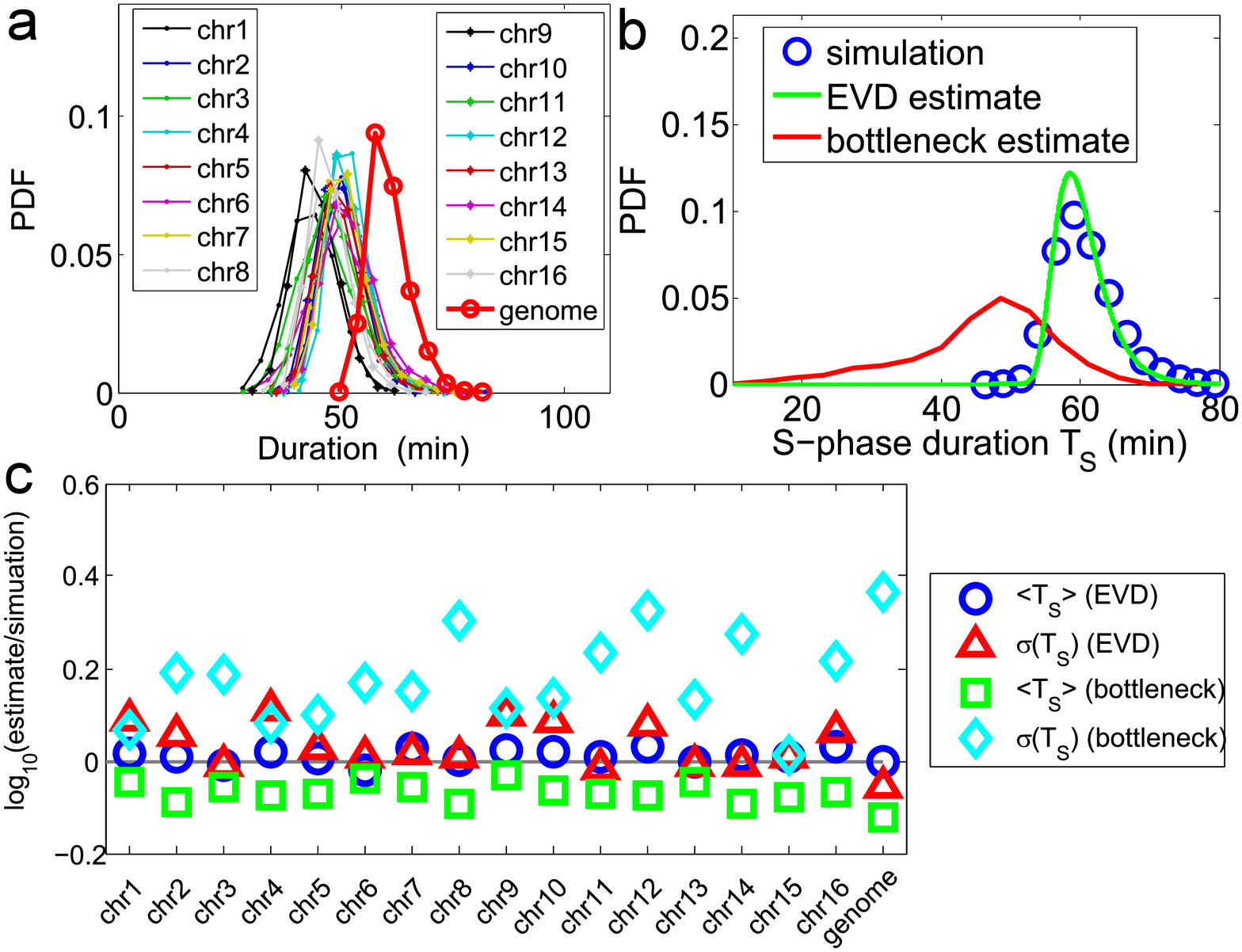}
 \caption{\textbf{Simulated and estimated prediction for the
     cell-to-cell variability of S-phase duration, using the
     best-fitting parameters for \emph{S.~cerevisiae}.} (a) The plot
   shows the probability density function (PDF) of the predicted
   duration of the replication of chromosomes and genome of
   \emph{S.~cerevisiae} from simulations. The average duration of S
   phase compares well with measurements from flow
   cytometry~\cite{hawkins2013high}. (b) The simulated distribution
   (PDF) of the replication timing of the genome (circle), is well
   predicted by EVD estimate (green line) rather than the bottleneck
   estimate (red line). (c) Comparison of the average and standard
   deviation of the duration of the replication from analytical
   estimates and the simulation. The EVD estimate predicts the
   replication timing of the genome and all the chromosomes better
   than the bottleneck estimate. Data from
   ref.~\cite{hawkins2013high}. chr: chromosome. }\label{fit2}
\end{figure}

\clearpage

\label{p-value}
\begin{figure}[h!]\centering
  \includegraphics[width=0.5\textwidth]{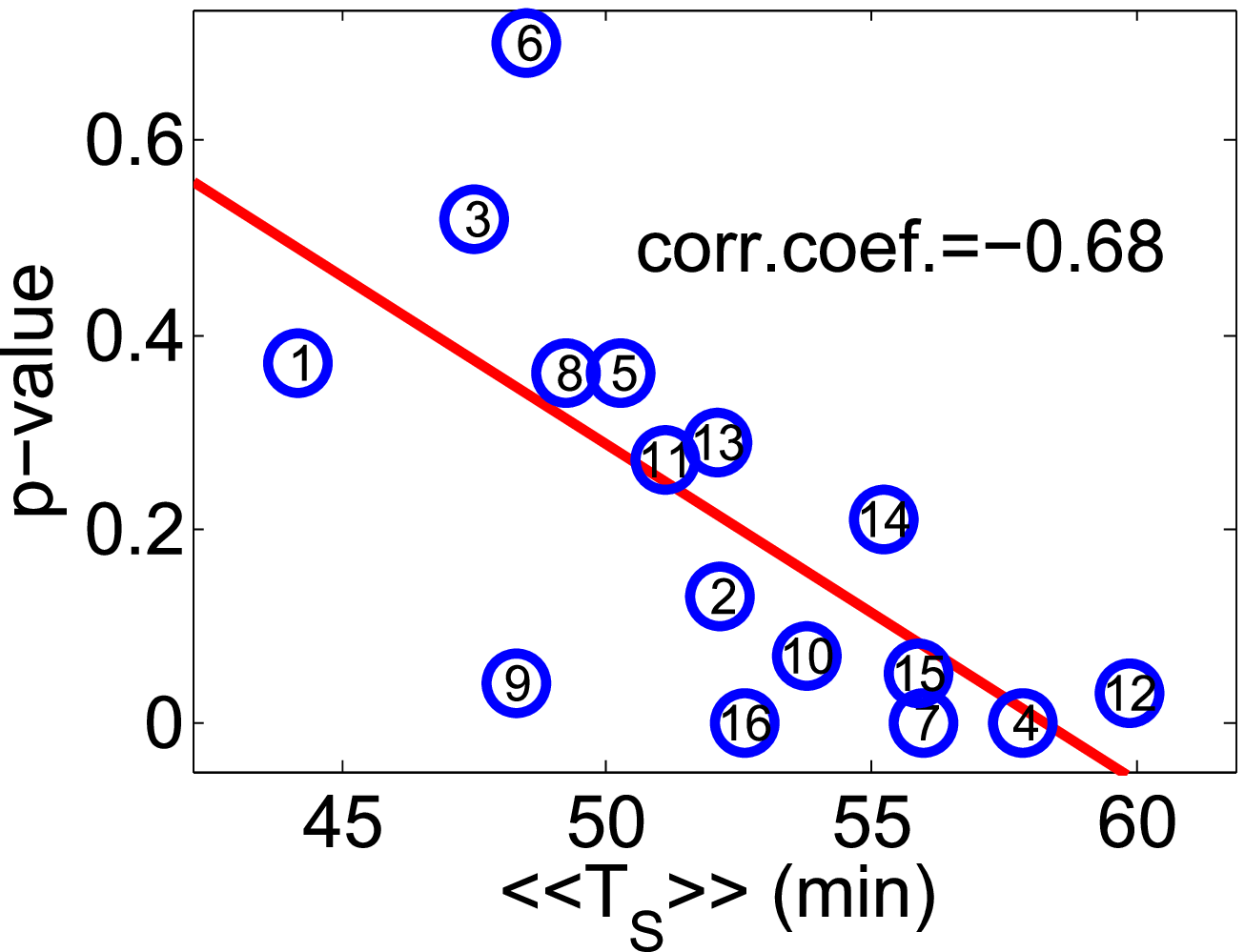}
  \caption{\textbf{Stronger bias towards smaller replication times for
      slower chromosomes. } The plot shows that the p-value of the
    mean $T_S$ for each \emph{S. cerevisiae} chromosome (circles,
    numbered 1-16) against randomized chromosomes is negatively
    correlated with the typical replication timing $T_S$ of the
    randomized chromosomes. See Fig.~5b of the main text.  Randomized
    chromosomes have the same averaged inter-origin distance and
    averaged origin strength. The typical time in the $x$-axis is
    defined as a double mean over realizations of the parameters and
    over cells, i.e., realizations of the process at fixed
    parameters. The P-value is defined as the fraction of the mean
    $T_S$ from randomized chromosomes smaller than the mean empirical
    $T_S$ over the number of randomised samples. }\label{p-value}
\end{figure}

\clearpage

\label{n_ori_chrom}
\begin{figure}[h!]\centering
  \includegraphics[width=1\textwidth]{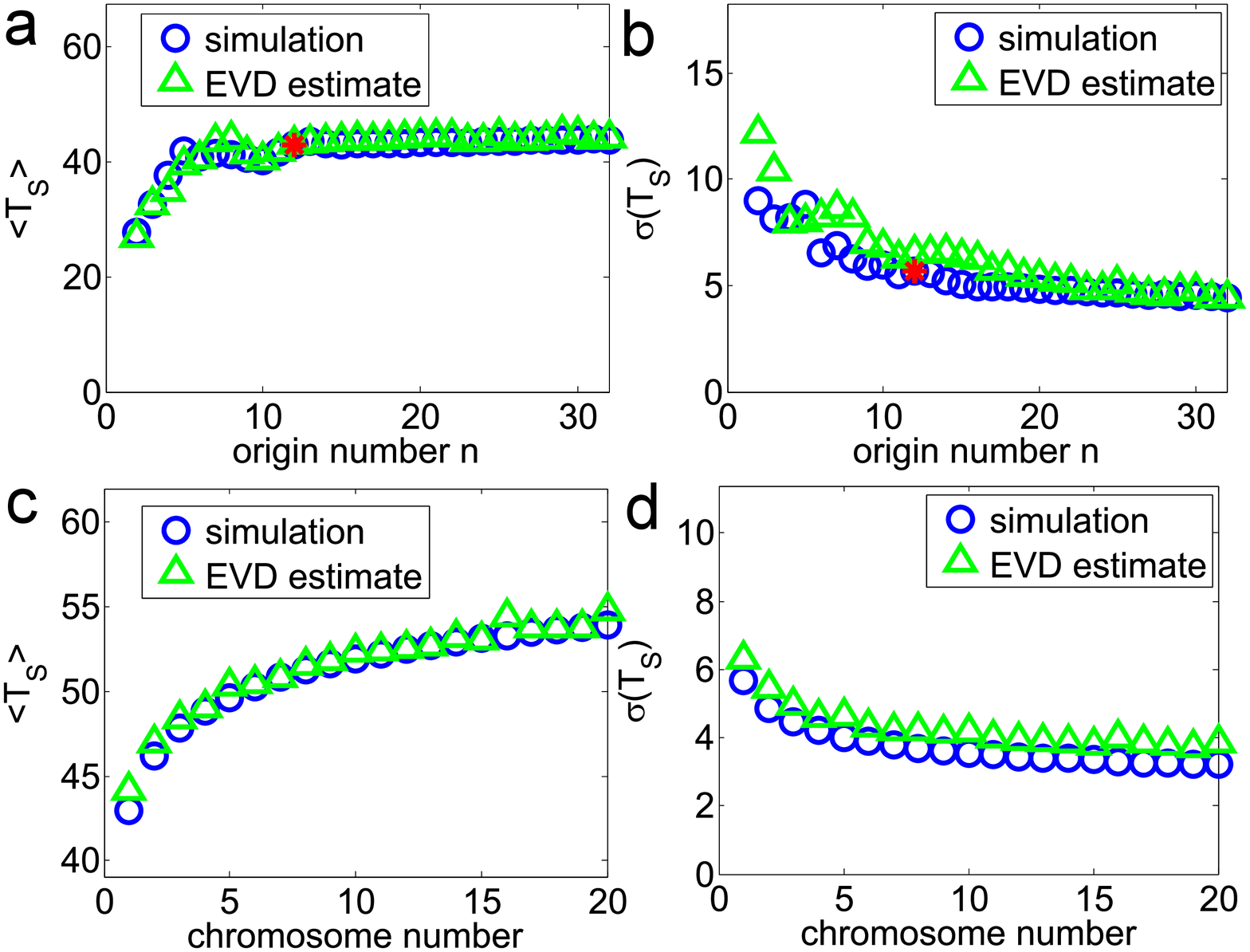}
  \caption{\textbf{Change of the overall replication timing and its
      cell-to-cell variability with number of inter-origin regions and
      with number of chromosomes.}  (a) Average of replication
    duration of \emph{S.~cerevisiae} chromosome I (parameters from the
    fit of data from ref.~\cite{hawkins2013high}) increases with
    origin number. The value saturates around n=10. (b) The standard
    deviation decreases with $n$. Red stars indicate the empirical
    value of $n$.
    (c,d) The average of the completion time for replication of
    \emph{S.cerevisiae} chromosome 1 increases with the number of
    copies, whereas the standard deviation decreases.
  }\label{n_ori_chrom}
\end{figure}


\clearpage

\label{Scerevisiae_originMutant}
\begin{figure}[h!]\centering
  \includegraphics[width=1\textwidth]{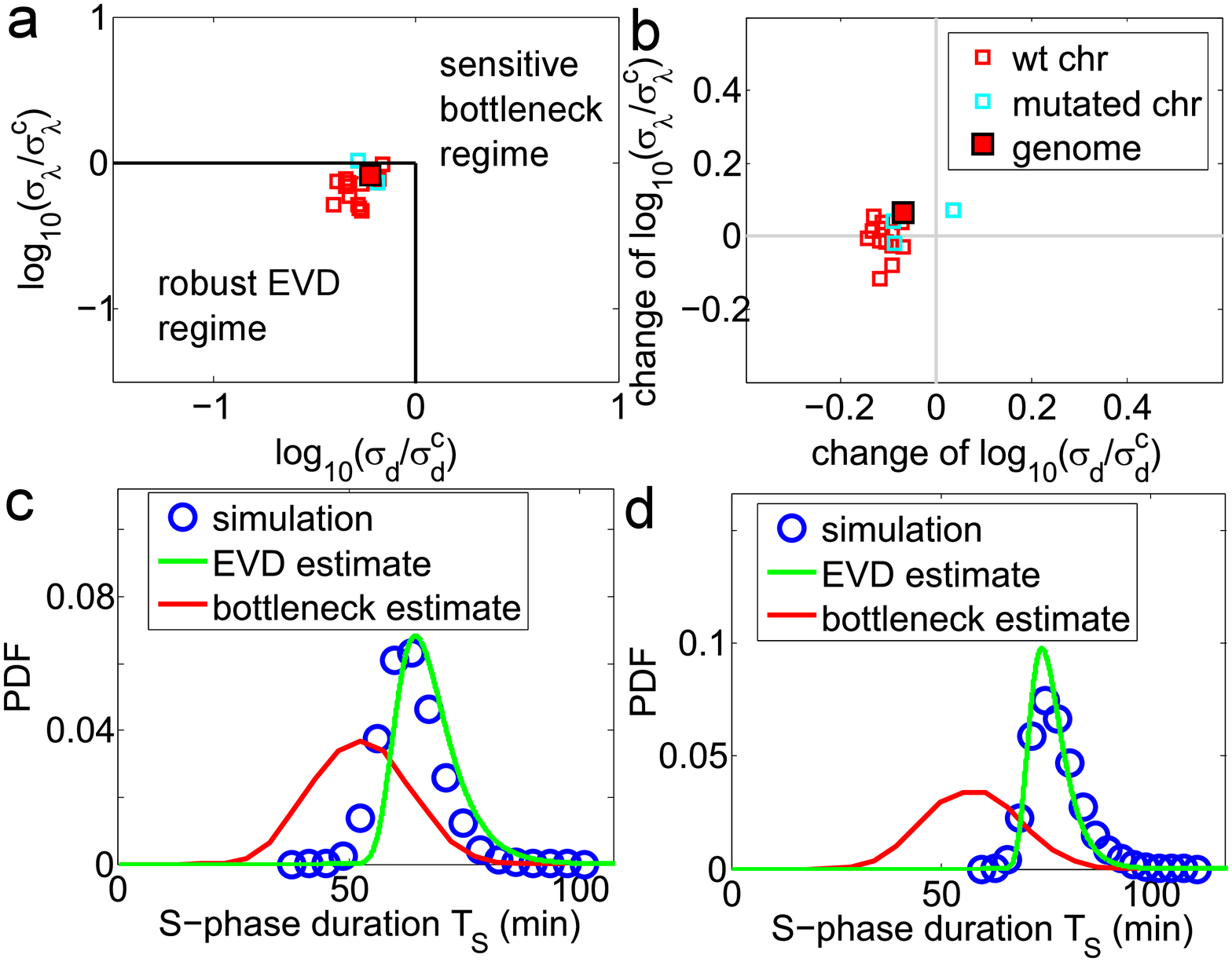}
  \caption{\textbf{ \emph{S.cerevisiae} remains in the extreme-value
      regime under inactivation of three origins in chromosomes 6, 7
      and 10.}  (a) The phase diagram indicates that all the
    chromosomes (except for chromosome 7) and the genome remain in the
    extreme-value regime when origins are removed.  (b) The overall
    relative variability of inter-origin distances for the mutant
    strain does not change much compared to the wt strain. The xy-axes
    indicate the change of the overall relative variability of
    inter-origin distances ($\log_{10}(\sigma_d/\sigma_d^c)$) and
    origin strengths ($\log_{10}(\sigma_\lambda/\sigma_\lambda^c)$) of
    the origin mutant strain compared to the wt strain. (c,d) The
    extreme-value estimate predicts well the replication duration of
    chromosomes (e.g. chromosome 7 shown in panel c) and the genome
    (panel d). The plots refer to fits of data of \emph{S.cerevisiae}
    origin-impaired mutant and wt strain from
    ref.~\cite{hawkins2013high}.}
  \label{Scerevisiae_originMutant}
\end{figure}

\clearpage

\label{Scerevisiae_originMutant_chr6}
\begin{figure}[h!]\centering
  \includegraphics[width=1\textwidth]{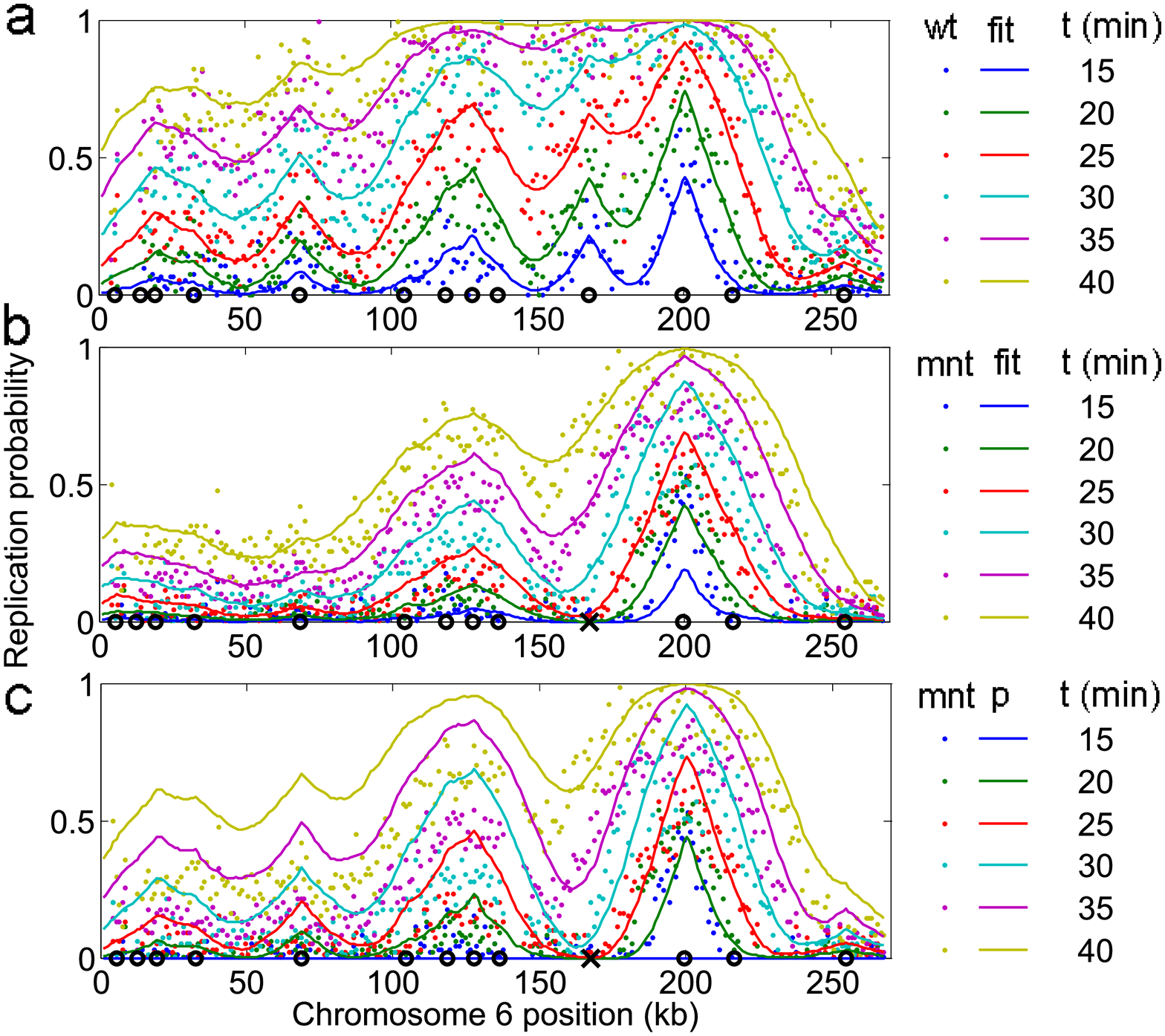}
  \caption{The model gives a satisfactory prediction of replication
    timing profiles of \emph{S.cerevisiae} origin mutant (mnt)
    strains. The plots refer to chromosome 6 as an example, and assess
    the performance of the model with parameters based on the
    wild-type fit on the mutant data, when simulations are run without
    the inactivated origins. Dots correspond to experimental data from
    ref.~\cite{hawkins2013high}, and lines indicate a model fit or a
    model prediction (p). Different dot colors correspond to different
    times. The black circles indicate origin locations. The black
    cross mark shows the location of the inactivated origin.  (a)
    Model fit of replication timing profiles of the wt strain. (b)
    Model fit of replication timing profiles of the origin mutant
    strain. (c) Model \emph{prediction} of mutant replication timing
    profiles based on the best-fit parameters from the wt data.}
  \label{Scerevisiae_originMutant_chr6}
\end{figure}

\clearpage

\label{Scerevisiae_Vincent}
\begin{figure}[h!]\centering
  \includegraphics[width=1\textwidth]{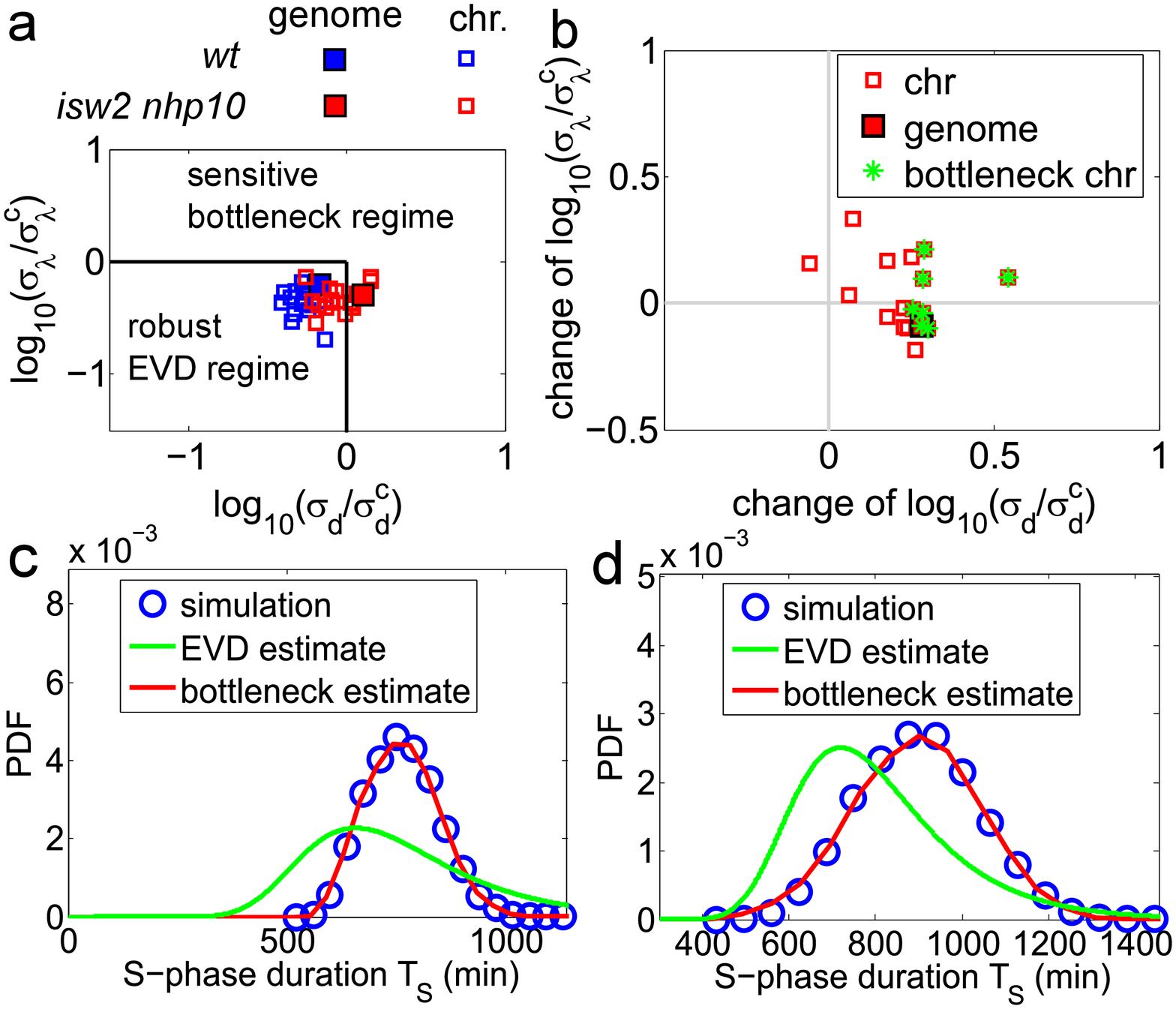}
  \caption{\textbf{ \emph{S.cerevisiae} \emph{isw2/nhp10} mutant
      treated with DNA alkylating agent MMS (affecting replication
      forks) drives S-phase to the bottleneck regime.} (a) The phase
    diagram (see Fig.~4 and 5 in the main text) indicates that all the
    chromosomes and the genome of the wt strain are in the EVD regime
    while some chromosomes (4, 6, 12, 13, 14 and 15) and the genome of
    the mutant are in the bottleneck regime (b) The relative
    variability of the inter-origin distances for the chromosomes and
    the genome of the mutant strain is higher that of the wt strain
    (except for chromosome 1). The green stars indicate that the
    chromosomes/genome of the mutant strain is inside the bottleneck
    regime. The xy-axes indicate the change of the overall relative
    variability of inter-origin distances
    ($\log_{10}(\sigma_d/\sigma_d^c)$) and origin strengths
    ($\log_{10}(\sigma_\lambda/\sigma_\lambda^c)$) of the
    \emph{isw2/nhp10} mutant strain compared to the wt strain. (c,d)
    The replication duration of some of \emph{S.cerevisiae}
    chromosomes, e.g. chr.~13 (shown in panel c) and 15 (panel d), in
    the mutant strain is well predicted by the bottleneck estimate
    rather than EVD estimate. Data of MMS (DNA alkylating agent methyl
    methanesulfonate) treated wild-type and \emph{isw2 nhp10} mutant
    strains of~\emph{S.cerevisiae} from
    ref.~\cite{vincent2008atp}. Origin locations are obtained from the
    literature \cite{hawkins2013high}. Origins with zero firing rate
    from the fit were deleted in the statistics on inter-origin
    distances and origin strengths.  }\label{Scerevisiae_Vincent}
\end{figure}

\clearpage

\label{Scerevisiae_Vincent_chr4}
\begin{figure}[h!]\centering
  \includegraphics[width=1\textwidth]{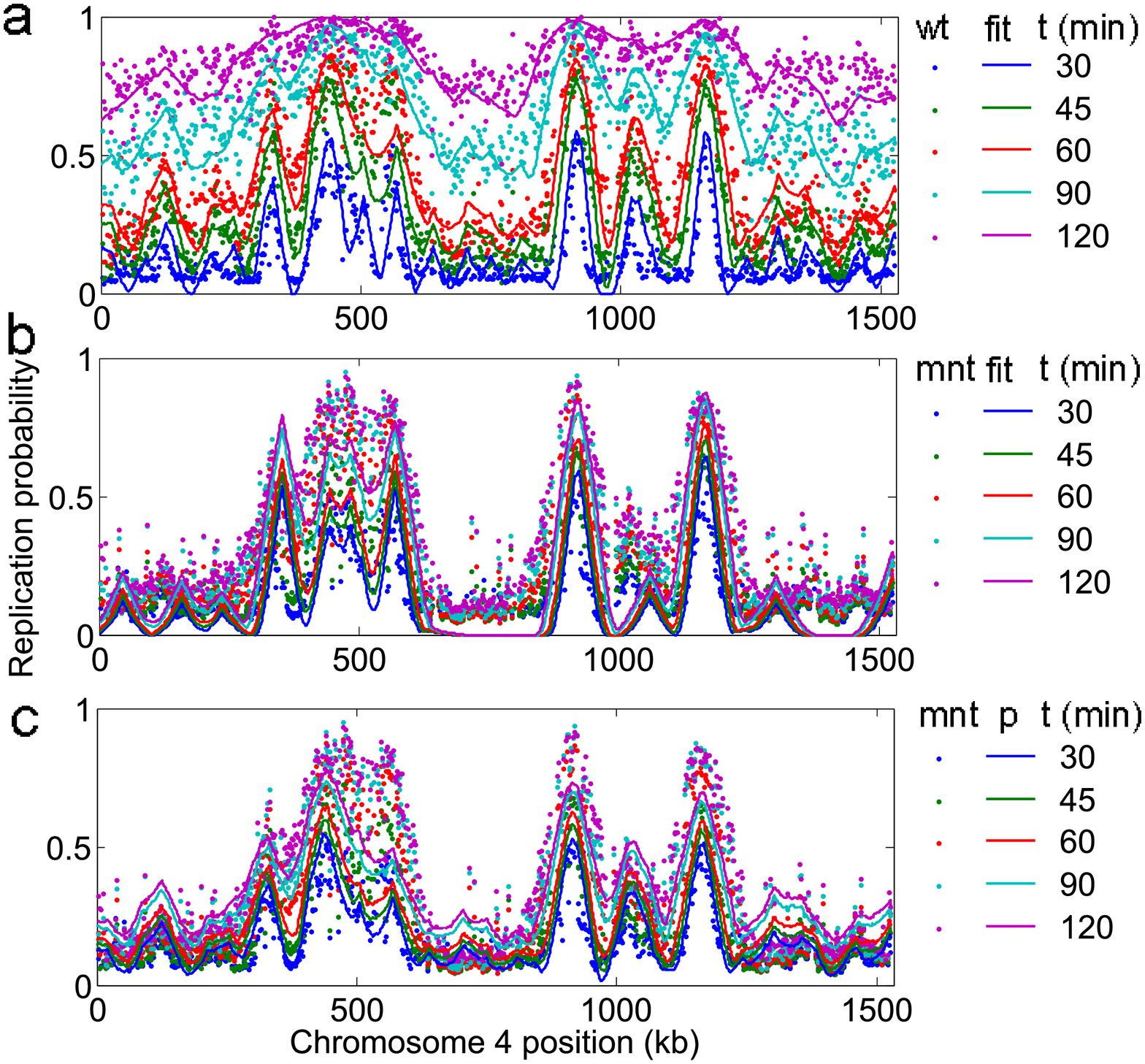}
  \caption{The model predicts well replication timing profiles of
    \emph{S.cerevisiae} \emph{isw2 nhp10} mutant (mnt) strains. The
    plots refer to chromosome 4 as an example. Dots correspond to
    experimental data of MMS-treated wild-type and \emph{isw2 nhp10}
    mutant strains of~\emph{S.cerevisiae} from
    ref.~\cite{vincent2008atp}, and lines indicates a model fit or
    prediction (p). Different dot colors correspond to different
    times.  (a) Model fit of replication timing profiles of chromosome
    4 of the wt strain. Origin locations from
    ref.~\cite{hawkins2013high} were used in this fit. (b) Model fit
    of replication timing profiles of chromosome 4 of the \emph{isw2
      nhp10} mutant strain. (c) Model prediction of mutant
    replication timing profiles based on the best-fit parameters from
    the wt data. The model parameters correspond to best-fit values of
    $\gamma$ and origin strengths from the wt data. For the
    prediction, all origin rates from the wt best fit were multiplied
    by an adjusted global constant factor (about 1/8), and fork speed
    and replication initial time were taken from the fit of mutant
    data.}
  \label{Scerevisiae_Vincent}
\end{figure}

\clearpage

\setcounter{table}{0}
\renewcommand{\thetable}{S\arabic{table}}

\begin{table}[h!]\begin{flushleft}
\caption{The parameters for genomes and chromosomes of \emph{S.cerevisiae}, \emph{L.kluyveri} and \emph{S.pombe} from the best fit
of genome-wide time-course replication data with the model. }
\label{table:01}%
 \medskip
{
\begin{tabular*}{0.55\columnwidth}{@{}lllllllll@{}}
\multicolumn{8}{c}{Parameters for \emph{S.cerevisiae (SC)}, \emph{L.kluyveri} and \emph{S.pombe} genomes } \\ \toprule
Species$^*$ & $\gamma^{\dag}$ & $v^{\dag}$  &$T_0^{\dag}$  & $\left<d\right>^\ddag$  & $\sigma_d^\ddag$ & $\left<\lambda\right>^\ddag$  & $\sigma_\lambda^\ddag$
\\
& & (kb/min) & (min) & (kb) &(kb) & ($\min^{-\gamma-1}$) & ($\min^{-\gamma-1}$)
\\
 \hline
\emph{SC} wt1 & 1.5 & 1.8 &1.3 & 26.1  & 16.9 & 5.3$\times 10^{-4}$ & 4.5$\times 10^{-4}$  &
\\
\emph{SC} mut1& 1.5 & 2.0 &5.0 & 26.2 & 17.2
 &1.9$\times 10^{-4}$ & 2.1$\times 10^{-4}$  &
 \\
\emph{SC }wt2  & 0.25 & 0.84 &-13 & 37.3 & 22.9
 &3.1$\times 10^{-3}$ & 2.0$\times 10^{-3}$  &
  \\
\emph{SC} mut2 & 0.75 & 0.27 &-161 & 85.4 & 64.3
 & 8.4$\times 10^{-5}$ & 5.6$\times 10^{-5}$  &
\\
\emph{L.kluyveri} & 1.75 & 2.5 &72.5 & 47.0  & 24.6 & 9.2$\times 10^{-5}$ & 6.2$\times 10^{-5}$ &
\\
\emph{S.pombe} & 2.0 & 2.55 &20.1 & 45.0  & 27.9 & 5.9 $\times 10^{-6}$ & 3.7$\times 10^{-6}$ &
\\
 \hline
\end{tabular*}%

 \bigskip

\begin{tabular*}{0.8\columnwidth}{@{}lllllllllllllllll@{}}
\multicolumn{17}{c}{Parameters for \emph{S.cerevisiae} wt1 chromosomes 1-16 (c1-c16)} \\ \toprule
Parameter & c1 & c2 & c3 & c4 & c5 & c6 & c7 & c8 & c9 & c10 & c11 & c12 & c13 & c14 & c15 &c16
\\
 \hline
 n$^\S$ & 12 & 34 & 15 & 51 & 20 & 13 & 41 & 26 & 20 & 26 & 24 & 40 & 35 & 25 & 40 & 37
 \\
$\left<d\right>$ (kb) & 19.3 & 23.6 & 20.0 & 30.0 & 28.1 & 20.8 & 26.2 & 21.3& 22.2 & 29.1 & 27.7 & 26.9 & 26.1 &31.3 & 27.0 & 25.5
\\
$\sigma_d$ (kb) & 12.8 & 13.6 & 13.2 & 20.0 & 13.5 & 12.3& 18.0& 14.8 & 16.4& 17.7 & 17.3 & 20.5 & 15.6& 16.0 &19.2& 15.0
\\
$\left<\lambda\right>$  & 4.5 & 4.3 & 5.0 & 6.0 & 6.2 & 4.7& 5.0& 4.8 & 4.9 & 5.4 & 5.7 &5.5 & 5.6& 5.3& 5.5 & 5.2
\\($\times10^{-4}\min^{-2.5}$) &&&& &&&&& & & & & & & &
\\
$\sigma_\lambda$   & 4.1 & 3.1 & 5.9 & 4.5 & 5.5 & 6.5& 4.5& 4.1& 3.4 & 4.5 & 4.3 & 6.0 & 4.3&4.4& 4.4& 3.8
\\($\times10^{-4}\min^{-2.5}$) &&&& &&&&& & & & & & & &
\\
 \hline
\end{tabular*}%


 \bigskip

\begin{tabular*}{0.8\columnwidth}{@{}lllllllllllllllll@{}}
\multicolumn{17}{c}{Parameters for \emph{S.cerevisiae} mut1 chromosomes 1-16 (c1-c16)}\\ \toprule
Parameter & c1 & c2 & c3 & c4 & c5 & c6 & c7 & c8& c9 & c10 & c11 & c12 & c13 & c14 & c15 &c16
\\
 \hline
 n & 12 & 34 & 15 & 51 & 20 & 12 & 40 & 26& 20 & 25 & 24 & 40 & 35 & 25 & 40 & 37
 \\														
$\left<d\right>$ (kb) &  19.4 & 23.6 & 20.0 & 29.7 & 28.1& 22.4 & 26.8 &21.1 & 22.2 & 30.1 & 27.7 & 26.8 & 26.0 &31.2 & 26.9& 25.5
\\
$\sigma_d$ (kb) & 12.7 & 13.7 & 13.3 & 20.0 & 13.5& 17.1& 19.2& 14.8 & 16.4 & 18.8 & 17.3 & 20.2 & 15.7& 16.0 &19.1& 15.0
\\
$\left<\lambda\right>$  & 1.7 & 1.5 & 2.8 & 2.0 & 2.7& 2.2& 1.8& 1.7& 2.2 & 2.2 & 1.7 &1.9 & 1.9& 1.7& 2.1 & 1.5
\\($\times10^{-4}\min^{-2.5}$) &&&& &&&&& & & & & & & &
\\
$\sigma_\lambda$ & 1.8 & 1.7 & 3.4 & 2.1 & 2.9 & 4.3 & 1.7& 1.5& 1.7 & 2.5 & 1.2 & 2.5 & 1.9&1.8& 2.1& 1.3
\\($\times10^{-4}\min^{-2.5}$) &&&& &&&&& & & & & & & &
\\
 \hline
\end{tabular*}%

%

 \bigskip

 \begin{tabular*}{0.8\columnwidth}{@{}lllllllllllllllll@{}}
\multicolumn{17}{c}{Parameters for \emph{S.cerevisiae} wt2 chromosomes 1-16 (c1-c16) treated with MMS}\\
 \toprule
Parameter & c1 & c2 & c3 & c4 & c5 & c6 & c7 & c8 & c9 & c10 & c11 & c12 & c13 & c14 & c15 &c16
\\
 \hline
 n & 6 & 20 & 10 & 36 & 17 & 7 & 32 & 15& 12 & 20 & 20 & 29 & 22 & 18 & 31 & 26
 \\										
$\left<d\right>$ (kb) &  41.0 & 40.6& 33.4& 41.7& 32.8& 41.3& 33.5& 35.6& 36.6& 37.4& 33.4& 36.5& 42.7& 42.6& 34.7& 35.4
\\
$\sigma_d$ (kb) &  40.4 & 26.1& 19.2& 25.4& 16.2& 35.3& 18.3& 20.2& 23.6& 21.8& 19.4& 25.8& 26.5& 19.1& 21.2& 22.8
\\
$\left<\lambda\right>$  & 3.4 & 3.0 & 3.6 & 2.7 & 3.3& 4.4& 3.0& 2.8 & 3.7 & 3.5 & 2.6 &3.2 & 3.6& 3.0& 2.9 & 2.8
\\($\times10^{-3}\min^{-1.25}$) &&&& &&&&& & & & & & & &   
\\
$\sigma_\lambda$ & 1.7 & 1.5 & 2.0 & 1.6 & 2.1 & 2.9 & 2.2& 1.3  & 3.0 & 2.9 & 2.0 & 1.7 & 1.8&2.1& 1.9& 1.7
\\($\times10^{-3}\min^{-1.25}$) &&&& &&&&& & & & & & & &   
\\
 \hline
\end{tabular*}%

\bigskip
 \begin{tabular*}{0.85\columnwidth}{@{}lllllllllllllllll@{}}
\multicolumn{17}{c}{Parameters for \emph{S.cerevisiae} mut2 chromosomes 1-16 (c1-c16) treated with MMS}\\ \toprule
Parameter & c1 & c2 & c3 & c4 & c5 & c6 & c7 & c8& c9 & c10 & c11 & c12 & c13 & c14 & c15 &c16
\\
 \hline
 n & 4 & 10 & 6 & 20 & 7 & 3 & 13 & 6& 4 & 9 & 6 & 14 & 9 & 5 & 15 & 13
 \\	
$\left<d\right>$ (kb) &  63.8& 80.6& 51.3& 75.5& 80.5& 119.0& 83.1& 93.3& 114.0& 87.9& 108.0& 78.5& 99.3& 166.0& 75.9& 71.2
\\
$\sigma_d$ (kb) &  51.5& 54.4& 28.7& 70.1& 42.6& 114.0& 50.4& 61.0&  63.4& 50.2&  58.7& 67.9& 79.4& 117.0& 67.5& 36.3
\\
$\left<\lambda\right>$  & 5.2& 9.2& 9.4& 5.2& 10.0& 12.2& 9.1& 7.9 & 16.4& 10.5& 8.9& 8.1& 10.1& 10.2& 6.9& 7.2
\\($\times10^{-5}\min^{-1.75}$) &&&& &&&&& & & & & & & &
\\
$\sigma_\lambda$ & 3.9& 6.1& 9.6& 4.7&  4.5&  8.7& 4.8& 5.0 & 5.1&  6.3& 4.2& 4.3&  5.7&  7.4& 4.6& 4.9
\\($\times10^{-5}\min^{-1.75}$) &&&& &&&&& & & & & & & &
\\
 \hline
\end{tabular*}%

}
{\textit{Continued on next page.}
}
\end{flushleft}
\end{table}

%
%

\clearpage
\setcounter{table}{0}
\begin{table}[h!]\begin{flushleft}
 \medskip
\caption{\textit{Continued from previous page} }

%

  \bigskip

 \begin{tabular*}{0.5\columnwidth}{@{}lllllllll@{}}
\multicolumn{9}{c}{Parameters for \emph{L.kluyveri} chromosomes 1-8 (c1-c8)} \\ \toprule
Parameter &c1 & c2 & c3 & c4 & c5 &c6 &c7 &c8
\\
 \hline
 n & 24 & 25 & 27& 27 & 30 & 31 & 43 & 39
 \\
$\left<d\right>$ (kb) & 42.6 & 44.9 & 46.9 & 49.1 & 43.9 & 44.9 & 41.3 & 59.8
\\
$\sigma_d$ (kb) & 29.0 & 22.6 & 18.0 & 24.4 & 14.1& 21.1& 22.0& 34.4
\\
$\left<\lambda\right>$ & 7.4 & 9.8 & 11.6 & 9.7 &7.0& 8.9&7.9 &11.4
 \\($\times10^{-5}\min^{-2.75}$) & & & & & & & &
\\
$\sigma_\lambda$  & 4.6 & 7.1 & 5.7 & 7.1 & 5.1& 5.3& 5.9& 6.8
\\($\times10^{-5}\min^{-2.75}$) & & & & & & & &
\\
 \hline
\end{tabular*}%

 \bigskip
\begin{tabular*}{0.45\columnwidth}{@{}llll@{}}
\multicolumn{4}{c}{Parameters for \emph{S.pombe} chromosomes 1-3 (c1-c3)} \\ \toprule
Parameter &c1 & c2 & c3
\\
 \hline
 n & 125 & 107 & 52
 \\
$\left<d\right>$ (kb) & 45.1 & 43.5 & 47.4
\\
$\sigma_d$ (kb) & 26.9 & 31.0 & 23.3
\\
$\left<\lambda\right>$ ($\times10^{-6}\min^{-3}$) & 5.5 & 5.0 & 8.5
\\
$\sigma_\lambda$ ($\times10^{-6}\min^{-3}$)  & 3.2 & 3.1 & 4.7
\\
 \hline
\end{tabular*}%
\\
{$^*$\emph{SC} wt1 and \emph{SC} mut1 are the wide-type and origin mutant strains of \emph{S.cerevisiae} respectively from Hawkins et al. \cite{hawkins2013high}. \emph{SC} wt2 and \emph{SC} mut2 are the wide-type and \emph{isw2nhp10} mutant strains of \emph{S.cerevisiae} respectively from vincent et al. \cite{vincent2008atp}. \\
$^\dag$ global parameters \\
$^\ddag$ statistics of local parameters (inter-origin distances and origin strengths). \\
$^\S$ origin numbers of \emph{L.kluyveri}, \emph{S.cerevisiae} and \emph{S.pombe} are from Agier et al.~\cite{agier2013spatiotemporal}, Hawkins et al.~\cite{hawkins2013high} and Heichinger et al.~\cite{heichinger2006genome} respectively. As for \emph{S.cerevisiae} origin mutant, three inactivated origins were deleted from the origin list. For \emph{S.cerevisiae} \emph{isw2nhp10} mutant, origins with zero strengths were removed. }
\end{flushleft}
 \end{table}

\end{document}